# Stability of quantum eigenstates and kinetics of wave function collapse in a fluctuating environment

Simone Chiarelli[1] and Piero Chiarelli[2,3]

*(1) Scuola Normale Superiore, Consoli del Mare, 1 , 56126, Pisa, Italy*

Email: simone.chiarelli@sns.it
Phone: +39-050-509686
Fax: +39-050.563513

*(2) National Council of Research of Italy, San Cataldo, Moruzzi 1, 56124, Pisa, Italy*

Email: pchiare@ifc.cnr.it.
Phone: +39-050-315-2359
Fax: +39-050-315-2166

*(3) Interdepartmental Center "E.Piaggio", Faculty of Engineering, University of Pisa, Diotisalvi 2, 56122, Pisa, Italy.*

Abstract: The work analyzes the stability of the quantum eigenstates when they are submitted to fluctuations by using the stochastic generalization of the Madelung quantum hydrodynamic approach. In the limit of sufficiently slow kinetics, the quantum eigenstates show to remain stationary configurations with a very small perturbation of their mass density distribution. The work shows that the stochastic quantum hydrodynamic model allows to obtain the definition of the quantum eigenstates without recurring to the measurement process or any reference to the classical mechanics, by identifying them from their intrinsic properties of stationarity and stability. By using the discrete approach, the path integral solution of the stochastic quantum-hydrodynamic equation has been derived in order to investigate how the final stationary configurations depend by the the initial condition of the quatum superposition of states.

The stochastic quantum hydrodynamics shows that the superposition of states can relax to different stationary states that, in the small noise limit, are the slightly perturbed quantum eigenstates. The work shows that the final stationary eigenstate depends by the initial configuration of the superposition of states and that possibly the probability transition to each eigenstates can satisfy the Born rule, allowing the decoherence process to be compatible with the Copenhagen interpretation of quantum mechanics.

# 1. Introduction

The unified approach to the quantum and the classical mechanics is one of the major goals of the nowadys physics [1-18].

This lack of unitary picture has lead to quantum outputs that contrast with our sense of reality [1-4]. A quantitative investigation of the problem was given by Bell [3] in response to the so called EPR paradox [2] where the quantum non-locality is analyzed respect to the notions of the classical freedom and the local relativistic causality.

The wave function, as 'the probability of finding a particle at some location, is the basic assumption of the Copenhagen interpretation of quantum mechanics [5–7, 12]. However, such approach leads to the counterintuitive conclusion that the physical state is just a probability wave until observed. The absence of the physical connetion with the pre-measure state fights against the common sense of reality and the independence of the real world by the observer [7]: The rigorous assumption of the probabilistic connection with the pre-measure world leads to consequence that the real state is not physically defined before the measure.

The situation is even more complicated by the fact that the process of the observation is out the Hamiltonian description of quantum mechanics. The need of performing a classical process, such as the measure, to define the quantum eigenstates, indeed, leads to a great theoretical loophole.

The unified theoretical connection between the quantum and the classical mechanics would explain how the laws of physics pass from the quantum behavior to the classical one: Many problems about how classical phenomena such as, measure, causality, locality, physical state of the external reality are compatible with the quantum mechanics and the wave-particle duality, will find their natural explanation.

On the other hand, if the wave function has a physically meaning, then, there must exist a defined mechanism (e.g., the so-called wave function collapse) describing the interaction with the observer and/or its evolution in the classical universe. In this case, there exists the problem about how the Schrödinger equation can be generalized [15, 16] to describe the irreversible quantum evolution [17-37].

In order to solve this problem, various interpretations of quantum mechanics like the many-worlds interpretation [19], the Bohmian mechanics [20-21], the modal interpretation [22], the relational interpretation [23], the consistent histories [24], the transactional interpretation [25, 26], the QBism [27], the Madelung quantum hydrodynamics [28-30], the decoherence approach [31] have been proposed.

The Madelung approach (that is a particular case of the Bohmian mechanics [32]) has both the important peculiarity to be mathematically equivalent to the Schrödinger one [Birula] and to treat the wavefunction evolution $\cdot = |\cdot| e^{i\frac{S}{\hbar}}$ in a classical-like way as the motion of the mass density $|\cdot|^2$ with the impluse $p_i = \frac{\partial S}{\partial q_i}$ under the effect of the quantum pseudo potential, introducing the concept of trajectories and the existence of physical reality independent by the measure.

The Madelung description disembogues into the classical mechanics as soon as $\hbar$ is set to zero for the macroscopic objects. For this reason it represents a good theoretical framework for describing the classical to quantum transition, mesoscale phenomena and the interaction between quantum and classical system [33-36].

For sake of truth, it must be observed that, even if the Planck's constant is small (respect to the scale of the problem), if we wipe it out (by hand), as well as the related quantum potential, from the quantum hydrodynamic equations, we also eliminate the stationary quantum eigenstates (see Appendix B) and deeply change the nature of the equations of motion. Thence, a well defined analytical procedure is needed to correctly pass from the quantum non-local description to the local classical one [37] where the Planck constant is quite small.

Others aspects of the open quantum evolution are captured by the decoherence approach that investigates the possibility of obtaining the classical state through the lost of quantum wave coherence generateed by the presence of the environment [38-41]. The decoherence is shown to be produced into the system by the fluctuations induced by the environment of the overall system. This approach has the weak side in the fact that, strictly speaking, the whole system is still quantum with superposition of states. This approach is not able to explain how the overall quantum system can lead to irreversible processes and how the observer can perform statistical measures (i.e., to be quantum de-coupled with the measured system) given that the quantum mechanics is reversible. To ovecome this problem, the relational quantum mechanics introduces the super-observer that is not entangled with the overall system [23]. Actually, this "ad hoc" postulate, is unsatisfactory and brings logical contraddictions.
From the experimental and numerical simulation point of view, there exist the evidence that the decoherence and the localization of quantum states can be generated by the interaction with the stochastic fluctuations of the environment [38,41].
Recently [42], the author has shown that, if we consider the not static picture of the universe, as suggested by De Sitter, with residual curvature oscillations and introduce the corresponding dark matter fluctuations, the Madelung description leads to the stochasitc quantum hydrodynamic model (SQHM) that in the limit of microscopic systems (whose physical length is much smaller than the De Broglie one) leads to the Langevin-Schrodinger equation. In the limit of zero fluctuations the SQHM redudes to the quantum mechanics that represents the deterministic limit of the stochasti theory. The stochastic quantum hydrodynamics show that, in the coarse grained large scale description, depending on the type of Hamiltonian potential, there exist the possibility that the quantum pseudo potential can be disregard leading to the emergence of the classical description.
In this work the authors analyze the stochastic quantum hydrodynamic problem with the help of the discrete path integral method.
The goal of the present work is:

1) The definition of the quantum eigenstates without any "empirical" reference to classical processes (that are out of the theory) but from properties describable by the theory itself (e.g.,. stability and stationarity);
2) To obtain the description of the wave-function collapse as (decoherent) relaxation toward a stationary state attributable to the deterministic eigenstates compatibly with the Copenaghen interpretation of quantum mechanics;
3) To check if the Born rule can be derived by repeated (uncorrelated) measurements on a single-quantum system in contact with the classical environment.

In the final section the authors analyze:

i. how the wave-particle duality evolves in the stochastic Madelung approach,
ii. the conditions under which the classical mechanics can be achieved on large scale systems,
iii. the interconnection between the quantum decoherence and the measurement process,
iv. the possible field of applications of the theory..

## 1.1 Thermal and quantum stochasticity

The similarity, between the stochastic process of the classical dynamics and those of the quantum mechanics, has attracted the attention of many researchers [43-45] and gained important results from the quantum path integral approach [46] that, as shown by Klinert [47], can be described as an immaginary-time Marcovian process.
In Appendix A, an introductory section has been devoted to put under light the similarities and the differences between the quantum immaginary-time stochastic process and the real-time stochastic process of the irreversible classical dissipative dynamics.

In order to describe the evolution of the quantum superposition of states in presence of noise, the conservation equation of the quantum mass density distribution [48] is analyzed in its stochastic version . .

## 2. The stochastic quantum hydrodynamic equation

As shown in refernce [42], in presence of vacuum dark mass fluctuations, the quantum-hydrodynamic equations of motion acquire the stochastic form

$$\dot{q}_i = \frac{p}{m}, \tag{1}$$

$$\dot{p}_j = -m©\dot{q}_{j(t)} - \frac{\partial\left(V_{(q)} + V_{qu(n)}\right)}{\partial q_j} + m´_{\ j(q,t,T)} \tag{2}$$

where

$$V_{qu(n)} = -\frac{\hbar^2}{2m}\frac{1}{n^{1/2}}\frac{\partial^2 n^{1/2}}{\partial q\ \partial q} \tag{3}$$

For sake of rigour, it must be noticed that the mass density $n$ is not the mass density of the quantum hydrodynamic description (i.e., $\tilde{n} = |\cdot\ |^2$) but it is the reduced spatial probability of the phase space mass density of the Smolukowski equation of the Marcovian process (2)[42] (below we refer to it as probability mass density (PMD). As shown in ref.[42], these variables are connected each other in the limit of small system velocity field $\dot{q}$ so that the difference between the force generated by the quantum potential of the two variables $n$ and $\tilde{n}$, can be approximated to read

$$\frac{\partial}{\partial q_{˜}}\left(V_{qu(n)} - V_{qu(\tilde{n})}\right) \cong m\bar{D}^{1/2}¶_{(t)} + A_{1(q,t)}\dot{q} + O(\dot{q}^2), \tag{4}$$

giving rise to a stochastic force noise with null mean and a drift term proportional to the velocity field $\dot{q}$. If $A_1$ is assumed constant, it equals the friction coefficient $m©$ [42] in (2).

Generaly speaking, the system of equations (1-3) are not able to describe the quantum system evolution under large fluctuations that may happen on very long time scale since, in this case, $\dot{q}$ can be very large (For more details, see ref.[42]).

For the sufficiently general case, to be of practical interest, the noise $´_{\ (q,t,T)}$ can be assumed Gaussian with null correlation time, isotropic into the space and uncorrelated among different co-ordinates such as [42]

$$<´_{\ j(q_{˜},t)}, ´_{\ i(q\ +ª,t+²)}> = <´_{\ j(q_{˜})}, ´_{\ i(q\ )}>_{(T)}\ F(ª)¢(²)¢_{˜}\ ¢_{ji} \tag{5}$$

where, for warrantig that the quantum mechanics is recovered in the deterministic limit, we assume that

$$\lim_{T\to 0}\ <´_{\ j(q_{˜})}, ´_{\ i(q\ )}>_{(T)} = 0 \tag{6}$$

where $T$ is the fluctuation-amplitude parameter (with the dimension of a temperature) and where, the shape of the correlation function $F(ª)$ reads [42]

$$\lim_{T\to 0} F(\text{a}) \propto \frac{1}{\text{a}_c} exp[-(\frac{\text{a}}{\text{a}_c})^2] \, . \qquad (7)$$

where

$$\text{a}_c = \sqrt{2}\frac{\hbar}{\sqrt{mkT}} \qquad (8)$$

is the De Broglie length.

Moreover, since near the deterministic quantum limit (i.e., for descriptions of system with physical length $\pounds$ such as $\frac{\pounds}{\text{a}_c} \ll 1$) the noise can be approximated by the form [42]

$$\text{´}_{(q,t,T)} = \text{©} D^{1/2} \text{¶}_{(t)} \qquad (9)$$

where [42]

$$\text{©} \cong \sim \frac{2kT}{mD} + O(D^{-2}) \qquad (10)$$

where $\sim$ is the numerical parameter that measures how the quantum hydrodynamic trajectories of motion are perturbed by fluctuations (leading to quantum decoherence and to energy dissipation). This parameter is specific for each considered system since the sensibility of the system to fluctuations is related to the characteristics of their trajectories of motion (e.g., Lyapunov exponents). This aspect goes beyond the purpose of this work and (10) is semi-empirically assumed here.

For $\sim = 0$ (as happens in the deterministic limit, see condition (18) below) the system maintains the quantum coherence with no dissipation.

For $\sim = 1$ we retrieve the Einstein relation

$$D = \frac{2kT}{m\text{©}} \qquad (11)$$

that holds for systems of monomolecular markovian classical particles such as the so called "dust matter" (the mass density, in form of "monomolecular dust", of the classical limit of the Madelung quantum hydrodynamics [48]).

Thence, for microscopic systems of physical length $\pounds$, such as $\pounds \ll \text{a}_c$, equation (1-2) give

$$\ddot{q}_{j(t)} = -\text{©}\dot{q}_{j(t)} - \frac{1}{m}\frac{\partial\left(V_{(q)} + V_{qu}\right)}{\partial q_j} + \text{©} D_{jm}^{1/2} \text{¶}_{m(t)}, \qquad (12)$$

that for problems where fast variables (with characteristic time $\text{²}_{ch} <\approx \frac{1}{\text{©}}$) are of no physical meaning, reduces to [49]

$$\dot{q}_{j(t)} = -\frac{1}{m\text{©}}\frac{\partial\left(V_{(q)} + V_{qu}\right)}{\partial q_j} + D_{jm}^{1/2} \text{¶}_{m(t)}, \quad . \qquad (13)$$

(This existence of stationary long-time asymptotical states is warranted by the dissipative force $-©\dot{q}_{(t)}$ that, for $V_{(q)}$ independent by time, leads to the stationary solution $lim_{t\to\infty} < \ddot{q}_{(t)} >= 0$ [49]).
In the case of unidimentional system treated in the present work, (13) simplifies to

$$\dot{q}_{(t)} = -\frac{1}{m©}\frac{\partial\left(V_{(q)}+V_{qu}\right)}{\partial q} + D^{1/2}¶_{(t)} \qquad (14)$$

By dimentional argument, $D$ and $©$ can be recast in the forms [42]

$$D = ¥_D\left(\frac{£}{\text{a}_c}\right)^p \frac{\hbar}{2m} \qquad (15)$$

$$© = 2\sim\frac{kT}{mD} = 4\sim\frac{kT}{¥_D\hbar}\left(\frac{\text{a}_c}{£}\right)^p \qquad (16)$$

leading to the SDE

$$\ddot{q}_{(t)} = -4\sim\frac{kT}{¥_D\hbar}\left(\frac{\text{a}_c}{£}\right)^p\dot{q}_{(t)} - \frac{1}{m}\frac{\partial\left(V_{(q)}+V_{qu}\right)}{\partial q} + 4\sim\frac{kT}{\hbar\sqrt{¥_D}}\left(\frac{\text{a}_c}{£}\right)^{p/2}\left(\frac{\hbar}{2m}\right)^{1/2}¶_{(t)} \qquad (17)$$

where the "dissipation parameter" $\sim$ , in the determinitic limit of quantum mechanics (i.e., for $T\to 0$) obeys to the condition

$$lim_{T\to 0}\sim = 0, \qquad (18)$$

For the description of sufficiently slow kinetics, equation (17) can be simplified to

$$\dot{q}_{(t)} = -\frac{¥_D\hbar}{4\sim mkT}\left(\frac{£}{\text{a}_c}\right)^p\frac{\partial\left(V_{(q)}+V_{qu}\right)}{\partial q} + \sqrt{¥_D}\left(\frac{£}{\text{a}_c}\right)^{p/2}\left(\frac{\hbar}{2m}\right)^{1/2}¶_{(t)}. \qquad (19)$$

that for instance, for p=2 ., by using the identities

$$D = \frac{2\sim kT}{m©} = ¥_D\left(\frac{£}{\text{a}_c}\right)^2\frac{\hbar}{2m} = ¥_D £^2\frac{kT}{4\hbar} \qquad (20)$$

$$© = 4\sim_{(T)}\frac{kT}{¥_D\hbar}\left(\frac{\text{a}_c}{£}\right)^2 = \frac{8\hbar\sim}{¥_D m£^2} \qquad 21)$$

leads to

$$\ddot{q}_{(t)} = -\frac{8\hbar\sim}{¥_D m£^2}\dot{q}_{(t)} - \frac{1}{m}\frac{\partial\left(V_{(q)}+V_{qu}\right)}{\partial q} + \frac{4\hbar\sim}{\sqrt{¥_D}m£}\sqrt{\frac{kT}{\hbar}}¶_{(t)} \qquad (22)$$

and for slow kinetics to

$$\dot{q}_{(t)} = -¥_D\frac{£^2}{8\hbar\sim}\frac{\partial\left(V_{(q)}+V_{qu}\right)}{\partial q} + \sqrt{¥_D}\frac{£}{2}\sqrt{\frac{kT}{\hbar}}¶_{(t)} \qquad (23)$$

Equation (23), for $m \approx 10^{-30}Kg$ and $£ \approx 10^{-10}m$ , can be used for kinetics with characteristic time

$$^{2}{}_{ch} \gg \frac{1}{©} = \frac{¥_D m £^2}{8\hbar \sim} \approx ¥_D \frac{10^{-17} s}{\sim}. \qquad (24)$$

## 2.1. The SQHM in fluctuating background

If we consider a single quantum system in a fluctuating vacuum whose curvature wringles generate mass density fluctuations $¢ n_{(q\sim)}$ with the variance $< ¢ n_{(q\sim)}, ¢ n_{(q\sim)} >$, it is possible to define the explicit form of the diffusion coefficient such as [42]

$$D = \frac{< {}^{'}{}_{(q\sim,t)}, {}^{'}{}_{(q\sim)} >}{©^2} = \frac{\hbar^4}{4m^4} \frac{1}{©^2} \frac{a^2}{a_c^{\,3}} < ¢ n_{(q\sim)}, ¢ n_{(q\sim)} >. \qquad (25)$$

where $\frac{a}{4®}$ is the interaction length of the potential $V_{(q)}$ (e.g., for the Lennard-Jones potential $a$ is the boson-boson s-wave scattering length). By introducing into (22) we obtain

$$¥_D = 2^8 \frac{£^2}{a^2} \frac{\sim^2_{(T)}}{< ¢ n_{(q\sim)}, ¢ n_{(q\ \ )} >_{(T)}} \sqrt{\frac{m}{kT}} \qquad (26)$$

that for $T = 1°K$ gives

$$¥_D \approx 7.7 \times 10^{-2} \frac{\sim^2}{< ¢ n_{(q\sim)}, ¢ n_{(q\ \ )} >}. \qquad (27)$$

Being $n \approx \frac{10^{-30} Kg}{\left(10^{-10} m\right)^3} \approx 1 \frac{Kg}{m^3}$ and by assuming, for instance at $T = 1°K$, that the mass density fluctuations $< ¢ n_{(q\sim)}, ¢ n_{(q\ \ )} >^{1/2} \approx \frac{1}{40c^2} \frac{\text{ev}}{\left(10^{-10} m\right)^3} \approx \frac{10^{-35} Kg}{\left(10^{-10} m\right)^3} \approx 10^{-5} \frac{Kg}{m^3}$, for $10^{-6} < \sim < 10^{-1}$, it follows that

$$10^3 < \frac{¥_D}{\sim} \approx 10^9 \sim < 10^8, \qquad (28)$$

and both

$$10^{-14} s < \frac{1}{©} < 10^{-9} s \qquad (29)$$

and

$$10^{-12} < D < 10^{-2} \qquad (30)$$

# 3. Integration of the quantum-stochasic motion equation

The Markovian process (23) for $V_{qu} = 0$, or $V_{qu} = const_{(q)}$, obeys to the Smoluchowski integro-differential equation for Markovian probability transition function (PTF) [50]

$$P(q,q_0 \,|\, t + {}^2, t_0) = \int_{-\infty}^{\infty} P(q,z \,|\, {}^2, t) P(z,q_0 \,|\, t - t_0, t_0) d^q z \qquad (31)$$

where the PTF $P(q,z \,|\, {}^2, t)$ represents the probability that an amount of the PMD $n(q,t)$ at time $t$, in a temporal interval ${}^2$, in a point $z$, transfers itself to the point $q$ [50].

Moreover, by expressing the conservation of the PMD $n$ in integral form, we can see that the PTF generates a displacement by a vector $(q,t) - (z,0)$ according to the rule [50]

$$n(q,t) = \int P(q,z\,|\,t,0)\, n_{(z,0)} d^r z \tag{32}$$

where $r$ is the number of dimensions of the representative space of the system.

### 3.1. The simplest case: The stationary quasi-eigenstates

Generally speaking, for the quantum case, equation (23) is not described by a Fokker-Planck equation (FPE), since the quantum potential $V_{qu_{(n)}}$ owns a functional dependence by $n(q,t)$ (and by the PTF $P(q,q_0|t+^2,t_0)$.

Nevertheless, if the initial distribution is stationary (e.g., a quantum eigenstates) and is close to the long-time final stationary distribution of the stochastic case (i.e., $n_{(q,t)} \approx n_{eq(q)}$ condition that can be checked at the end) it is possible to assume the approximation

$$V_{qu} \cong -(\frac{\hbar^2}{4m})\left(\left(\frac{\partial^2 \ln n_{eq(q)}}{\partial q^2}\right) + \frac{1}{2}\left(\frac{\partial \ln n_{eq(q)}}{\partial q}\right)^2\right) \tag{33}$$

where the final stationary PMD $n_{eq}$ is independent by time (and by the system evolution) so that, close to the stationary condition, (23) can be approximately described by the Fokker-Plank equation

$$\frac{\partial P_{(q,z|t,0)}}{\partial t} + \frac{\partial P_{(q,z|t,0)}{}^{3}}{\partial q} = 0 \tag{34}$$

where

$$^{3} = -\frac{1}{m\copyright}\frac{\partial\left(V_{(q)} - (\frac{\hbar^2}{4m})\left(\left(\frac{\partial^2 \ln n}{\partial q^2}\right) + \frac{1}{2}\left(\frac{\partial \ln n}{\partial q}\right)^2\right)\right)}{\partial q} - \frac{D}{2}\frac{\partial \ln n}{\partial q} \tag{35}$$

leading to the final equilibrium identity

$$\frac{1}{m\copyright}\frac{\partial\left(V_{(q)} - (\frac{\hbar^2}{4m})\left(\left(\frac{\partial^2 \ln n_{eq}}{\partial q^2}\right) + \frac{1}{2}\left(\frac{\partial \ln n_{eq}}{\partial q}\right)^2\right)\right)}{\partial q} + \frac{D}{2}\frac{\partial \ln n_{eq}}{\partial q} = 0 \tag{36}$$

### 3.1.2 Linear systems

In the case of linear systems

$$V_{(q)} = \frac{m\mu^2}{2} q^2 ,\qquad (37)$$

the equlibrium condition leads to

$$\frac{\hbar^2}{4m} \frac{\partial \left( \left( \frac{\partial^2 ln\, n}{\partial q^2} \right) + \frac{1}{2} \left( \frac{\partial\, ln\, n}{\partial q} \right)^2 \right)}{\partial q} = m\copyright D \frac{\partial\, ln\, n}{\partial q} + m\mu^2 q \qquad (38)$$

that is satisfacted by the solution $n = n_0\, exp\left[ -\frac{q^2}{\Delta q^2} \right]$, where $n_{0(q)}$ is defined by the relation

$$\frac{\hbar^2}{4m^2} \frac{\partial \left( \left( \frac{\partial^2}{\partial q^2} \left( ln\, n_0 - \frac{q^2}{\Delta q^2} \right) \right) + \frac{1}{2} \left( \frac{\partial}{\partial q} \left( ln\, n_0 - \frac{q^2}{\Delta q^2} \right) \right)^2 \right)}{\partial q} = \copyright D \frac{\partial \left( ln\, n_0 - \frac{q^2}{\Delta q^2} \right)}{\partial q} + \mu^2 q \qquad (39)$$

For a stationary state owing the initial PMD of the foundamental state $n^{1/2} = |\cdot| = C_0\, exp\left[ -\frac{q^2}{2\Delta q^2} \right]$ where $C_0 = \left( \frac{1}{4\circledR} \frac{m\mu}{\hbar} \right)^{1/4}$, from (39) it follows that

$$\frac{1}{\Delta q^2} = -\frac{m^2}{\hbar^2} \copyright D + \frac{m\mu}{\hbar} \sqrt{1 + \left( \frac{m}{\hbar\mu} \copyright D \right)^2} \qquad (40)$$

that near the quantum mechanics (i.e., $D \to 0$, $\copyright \to 0$ and $\mu \gg \frac{m}{\hbar} \copyright D$) leads to

$$\frac{1}{\Delta q^2} \cong \frac{m\mu}{\hbar} - \frac{m^2}{\hbar^2} \copyright D + \frac{1}{2} \frac{m^3}{\hbar^3 \mu} (\copyright D)^2 \cong \frac{m\mu}{\hbar} - \frac{m^2}{\hbar^2} \copyright D \cong \frac{m\mu}{\hbar} \left( 1 - 2\sim \frac{kT}{\hbar\mu} \right). \qquad (41)$$

and to the distribution

$$n_{eq} = C\, exp\left[ -\frac{m\mu}{\hbar} \left( 1 - 2\sim \frac{kT}{\hbar\mu} \right) q^2 \right] \qquad (42)$$

Where $C$ is defined by the normalization condition.

In the limit of small thermal fluctuations the mass density distribution of the ground eigenstate (42) does not lose its Gaussian shape and acquires a little enlargement of its variance

$$\Delta q^2 = \frac{1}{\left( \frac{m\mu}{\hbar} - \frac{2\sim mkT}{\hbar^2} \right)} \cong \frac{\hbar}{m\mu} \left( 1 + 2\sim \frac{kT}{\hbar\mu} \right) = \Delta q_0^2 \left( 1 + \sim \frac{kT}{\hbar\mu} \right) \quad . \qquad (43)$$

where, for $\mathcal{L} \ll a_c$ (near the quantum limit) $\sim \ll 1$ and where

$$\Delta q_0{}^2 = \left( \frac{\hbar}{m\mu} \right) \quad . \qquad (44)$$

is the variance of the deterministic quantum state. As final check, we can see the result (42) satisfies the condition (33).

Besides, the energy $E_0$ of the fundamental stationary state, in presence of fluctuation, reads

$$
\begin{aligned}
E_0 =< \cdot_0 | H | \cdot_0 >= & \int_{-\infty}^{\infty} n_{(q,t)} \left[ \frac{m}{2}\dot{q}^2 + \frac{m\mu^2}{2} q^2 + V_{qu} \right] dq \\
= & \int_{-\infty}^{\infty} n_{(q,t)} \left[ \frac{m\mu^2}{2} q^2 + V_{qu} \right] dq \\
= & \int_{-\infty}^{\infty} n_{(q,t)} \left[ \frac{m\mu^2}{2} q^2 - \frac{m\mu^2}{2} \left( 1 - 2\sim \frac{kT}{\hbar\mu} \right) q^2 + (n + \frac{1}{2}) \hbar\mu \right] dq \\
= & \int_{-\infty}^{\infty} n_{(q,t)} \left[ \sim \frac{\mu m k T}{\hbar} q^2 + \frac{\hbar\mu}{2} \right] dq \\
= & \frac{\hbar\mu}{2} + \sim \frac{\mu m k T}{\hbar} \int_{-\infty}^{\infty} q^2 \left( \frac{m\mu}{\circledR \hbar} \right)^{1/2} exp\left[ - \left( \frac{m\mu}{\hbar} - \frac{2\sim m k T}{\hbar^2} \right) q^2 \right] dq \\
= & \frac{\hbar\mu}{2} + \sim \frac{\mu m k T}{\hbar} \left( \frac{1}{1 + 2\sim \frac{kT}{\hbar\mu}} \right) \Delta q^2 \\
\cong & \frac{\hbar\mu}{2} + \sim kT
\end{aligned} \quad , (45)
$$

showing an energy increases of $\sim kT$.

As far as it concerns the energy variance of the fundamental stationary state in presence of fluctuation

$$\Delta E_0 =< \cdot_0 | \left( E - E_0 \right)^2 | \cdot_0 >^{1/2}, \qquad (46)$$

it reads

$$
\begin{aligned}
\left( \Delta E_0 \right)^2 = & \int_{-\infty}^{\infty} n_{(q,t)} \left[ \left( \frac{m}{2}\dot{q}^2 + \frac{m\mu^2}{2} q^2 + V_{qu} \right) - \left( \frac{\hbar\mu}{2} + \sim kT \right) \right]^2 dq \\
= & \int_{-\infty}^{\infty} n_{(q,t)} \left[ \sim \frac{\mu m k T}{\hbar} q^2 - \sim kT \right]^2 dq \\
= & \left( \sim kT \right)^2 \int_{-\infty}^{\infty} n_{(q,t)} \left[ \frac{\mu m}{\hbar} q^2 - 1 \right]^2 dq = \sim kT \left( -1 + \frac{\mu m}{\hbar} \int_{-\infty}^{\infty} n_{(q,t)} \left( 2q^2 + \frac{\mu m}{\hbar} q^4 \right) dq \right) \\
\cong & \left( \sim kT \right)^2 \left( -1 + 1 + \left( \frac{\mu m}{\hbar} \right)^2 \int_{-\infty}^{\infty} n_{(q,t)} q^4 dq \right) \cong \frac{3}{8} \left( \sim kT \right)^2
\end{aligned} \qquad (47)
$$

that allows to derive the dissipation parameter $\sim$ from the energy variance of the foundamental state such as

$$\sim \cong \sqrt{\frac{8}{3}} \frac{\Delta E_0}{kT} \qquad (48)$$

### 3.1.2.1 Asymptotic stationary states (quasi-eigenstates) of harmonic oscillator in presence of noise

As shown in appendix C, the mass density distribution of the j-th excited eigenstates in presence of noise, at long-time equilibrium, reads

$$n^{1/2}{}_j = \bar{C}_j H_{j\left(q\sqrt{\frac{m\mu}{\hbar}}\right)} exp\left[ -\frac{1}{2}\left( \frac{m\mu}{\hbar} - \left( 1+\frac{\Delta q_0{}^2}{q^2} f_{j(q)} \right)\frac{2\sim mkT}{\hbar^2} \right) q^2 \right]$$
$$\cong \bar{C}_j\, exp\left[ -f_{j(q)}\frac{\sim kT}{\hbar\mu} \right] H_{j\left(q\sqrt{\frac{m\mu}{\hbar}}\right)} exp\left[ -\frac{1}{2}\frac{m\mu}{\hbar}\left( 1-\frac{2\sim kT}{\hbar\mu} \right) q^2 \right] \quad (49)$$

where $\bar{C}_j$ is the normalization constant, with $kT \ll \hbar\mu$ , $0 < \sim \ll 1$ .and where $f_{j(q)}$ obeys to the differential equation

$$H_j\left( \frac{\partial^2}{\partial q^2} f_{j(q)} \right) + \left( \frac{\partial}{\partial q} f_{j(q)} \right)\left( 4j\frac{H_{j-1}}{\Delta q_0} - 2\frac{qH_j}{\Delta q_0^2} \right) - \frac{8H_j\, ln H_j}{\Delta q_0{}^2} - \left( \frac{q}{\Delta q_0^2} 4jH_{j-1} \right) = 0 \quad (50)$$

that is the Bernoulli equation for $f'_{j(q)} = \frac{\partial}{\partial q} f_{j(q)}$ , where $H_j$ is the j-th Hermite polynomial ,leading to the solution

$$f_{j(q)} = \int_0^q \frac{ \int_0^x \left( \frac{8H_{j(u)}\, ln H_{j(u)}}{\Delta q_0{}^2} - \left( 4jx\frac{H_{j-1(u)}}{\Delta q_0^3} \right) \right) exp[ \int_0^u \left( 4j\frac{H_{j-1(t)}}{\Delta q_0} - 2t\frac{H_{j(t)}}{\Delta q_0^2} \right) dt ]\, du }{ exp[ \int_0^x \left( 4j\frac{H_{j-1(u)}}{\Delta q_0} - 2x\frac{H_{j(u)}}{\Delta q_0^2} \right) du ] } dx \quad (51)$$

From expression (49) we can see that, in the presence of thermal fluctuations, the eigenstates maintain their (deterministic) shape (C.? in appendix C) with a small increase of their variance: In this sense we can say that they are stable.
It is noteworthy to see that the parameter $\sim$ can be experimentally evaluated by the measure both of (48) or by the measure of the energy gap $\Delta E_j = E_j - E_{j-1}$ between eigenstate, where $E_j$ that can be obtained by the formula

$$E_j = < \cdot_j | H | \cdot_j > = \int_{-\infty}^{\infty} n_{(x,t)} \left[ \frac{m\mu^2}{2} q^2 - \frac{m\mu^2}{2} \left( 1 - \left( 1 + \frac{2}{q^2} \frac{\hbar}{m\mu} f_{j(q)} \right) 2 \sim \frac{kT}{\hbar\mu} \right) q^2 + (n + \frac{1}{2}) \hbar\mu \right] dq$$

$$= \int_{-\infty}^{\infty} n_{(x,t)} \left[ - \left( 1 + \frac{2}{q^2} \frac{\hbar}{m\mu} f_{j(q)} \right) \sim \frac{\mu m k T}{\hbar} q^2 + (n + \frac{1}{2}) \hbar\mu \right] dq$$

$$= (j + \frac{1}{2}) \hbar\mu - \sim \frac{\mu m k T}{\hbar} \bar{C}_j^{\,2} \int_{-\infty}^{\infty} \begin{pmatrix} \left( 1 + \frac{2}{q^2} \frac{\hbar}{m\mu} f_{j(q)} \right) q^2 H_j^{\,2} \, exp\left[ - f_{j(q)} \frac{2 \sim kT}{\hbar\mu} \right] \\ exp\left[ - \frac{m\mu}{\hbar} \left( 1 - \frac{2 \sim kT}{\hbar\mu} \right) q^2 \right] \end{pmatrix} dq$$

$$= (j + \frac{1}{2}) \hbar\mu - \left( 1 + \Delta_j \right) \sim \frac{\mu m k T}{\hbar} \Delta q^2$$

$$\cong (j + \frac{1}{2}) \hbar\mu - \left( 1 + \Delta_j \right) \sim kT$$

(52).

where $\Delta q^2 = \frac{1}{\frac{m\mu}{\hbar} \left( 1 - \frac{2 \sim kT}{\hbar\mu} \right)}$, where $\bar{C}_j$ is the normalization constant of the $n$-th eigenfunction,

and where

$$\Delta_j = \frac{\bar{C}_j^{\,2} \int_{-\infty}^{\infty} \begin{pmatrix} \left( 1 + \frac{2}{q^2} \frac{\hbar}{m\mu} f_{j(q)} \right) q^2 H_n^{\,2} \, exp\left[ - f_{j(q)} \frac{2 \sim kT}{\hbar\mu} \right] \\ exp\left[ - \frac{m\mu}{\hbar} \left( 1 - \frac{2 \sim kT}{\hbar\mu} \right) q^2 \right] \end{pmatrix} dq}{\bar{C}_0^{\,2} \int_{-\infty}^{\infty} q^2 \, exp\left[ - \frac{m\mu}{\hbar} \left( 1 - \frac{2 \sim kT}{\hbar\mu} \right) q^2 \right] dq} - 1 \cong \frac{\Delta q_j^2}{\Delta q^2} - 1 . \quad (53)$$

where the wideness of the $j$-th eigenstate reads

$$\Delta q_j^2 = \bar{C}_j^{\,2} \int_{-\infty}^{\infty} \begin{pmatrix} \left( 1 + \frac{2}{q^2} \frac{\hbar}{m\mu} f_{j(q)} \right) q^2 H_n^{\,2} \, exp\left[ - f_{j(q)} \frac{2 \sim kT}{\hbar\mu} \right] \\ exp\left[ - \frac{m\mu}{\hbar} \left( 1 - \frac{2 \sim kT}{\hbar\mu} \right) q^2 \right] \end{pmatrix} dq . \quad (54)$$

Finally, by using (52-53), it follows that

$$\sim \cong \frac{\Delta q^2}{\left( \Delta q_j^2 - \Delta q_{j-1}^2 \right)} \frac{\hbar\mu - \left( E_{j+1} - E_j \right)}{kT} \quad (55)$$

.

## 3.2. Evolution of quantum superposition of states submitted to fluctuations

In order to determine the evolution of quantum superposition of states, whose initial condition is not stationary, we need to integrate the stochastic differential equation (SDE)

$$\dot{q} = -¥_D \frac{£^2}{8\sim\hbar} \frac{\partial \left( V_{(q)} - (\frac{\hbar^2}{2m}) \left( \left( \frac{\partial^2 ln\, n}{\partial q^2} \right) + \left( \frac{\partial ln\, n}{\partial q} \right)^2 \right) \right)}{\partial q} + \sqrt{¥_D} \frac{£}{2} \sqrt{\frac{kT}{\hbar}} ¶_{(t)} \quad (56$$

As shown below, this can be done by using the discrcete approach with the help of the Smolukowski integro-differential equation (31) and the associated conservation equation (32) for the PMD.

Given the presence of the quantum pseudo-potential, this procedure is not general and holds under some approximations (see condition (63) below ).

We integrate the SDE (56) by using its 2nd order discrete expansion

$$q_{k+1} \cong q_k - \frac{1}{m©} \frac{\partial \left( V_{(q_k)} + V_{qu(n_{q_k})} \right)}{\partial q_k} \Delta t_k - \frac{1}{m©} \frac{d}{dt} \frac{\partial \left( V_{(q_k)} + V_{qu(n_{q_k})} \right)}{\partial q_k} \frac{\Delta t_k{}^2}{2} + D^{1/2} \Delta \mathcal{W}_k \quad (57)$$

where

$$q_k = q_{(t_k)} \quad (58)$$

$$\Delta t_k = t_{k+1} - t_k \quad (59)$$

$$\Delta \mathcal{W}_k = \mathcal{W}_{(t_{k+1})} - \mathcal{W}_{(t_k)} \quad (60)$$

where the PMD $n(q_k, t_k)$ from (32) reads

$$n(q_k, t_k) = \int_{-\infty}^{\infty} P(q_k, q_0 \,|\, t_k, 0) n(q_0, 0) dq_0 \quad (61)$$

where $\Delta \mathcal{W}_k$ has Gaussian zero mean and unitary variance probability function $\mathcal{P}(\Delta \mathcal{W}_k, \Delta t)$ that ,for $\Delta t_k = \Delta t \quad \forall\ k$ , reads

$$\mathcal{P}(\Delta \mathcal{W}_k, \Delta t) = (4® D \Delta t)^{-1/2} exp - \frac{1}{4} \frac{(\Delta \mathcal{W}_k)^2}{\Delta t} . \quad (62)$$

Before, proceding with the integration of the discrete motion equation, we observe that, generally speaking, given the link between the quantum potential and the PMD $n(q_k, t_k)$, the evolution of equation (57) depends by the sequence of the noise inputs $D^{1/2} \Delta \mathcal{W}_k$ and, thence, by the discrete time interval of integration.

This behaviour can be easily checked by making a numerical Montecarlo integration of (57) by using a stochastic routine generating the noise $D^{1/2} \Delta \mathcal{W}_k$ .

In order to solve this vagueness it is necessary to postulate that the physical time is not continuous but discrete (this can be done by assuming the Planck time $t_p = \sqrt{\frac{\hbar G}{c^5}}$ as the ultimate physical time interval) and, for systems whose kinetics evolve on a time scale, usually by far much larger than $t_p$ , that the discrete time interval $\Delta t$ satisfies the condition

$$< \Delta \mathcal{W}_k, \Delta \mathcal{W}_k >^{1/2} = \sqrt{D \Delta t} \ll \left| \frac{1}{m©} \left( \frac{\partial \left( V_{(q_k)} + V_{qu(n)} \right)}{\partial q_k} \Delta t - \frac{d}{dt} \frac{\partial \left( V_{(q_k)} + V_{qu(n)} \right)}{\partial q_k} \frac{\Delta t^2}{2} \right) \right. \quad (63).$$

Condition (63) requires that, the integration time interval must be choosen as short as to have the force noise variance sufficiently small respect to the determiistic force (but also sufficiently large to warrant condition (25) $\Delta t \gg \frac{1}{\text{©}} = \frac{m\mathcal{L}^2}{8\text{¥}_D \hbar\text{~}}$.

These conditions can be both satisfied for $D \to 0$.

Under these approximations, the PTF solution reads (see appendix D)

$$P(q,q_0,t,t_0)=$$

$$= lim_{\Delta t\to 0}\left\{\Pi_{k=1}^{n}\int_{-\infty}^{\infty} dq_{k-1}\right\}\left(4\text{®} D\Delta t\right)^{-n/2} exp\left[\begin{array}{l} -\frac{1}{2D}\sum_{k=1}^{n} < \dot{\bar{q}}_{k-1} > \Delta q_k - \frac{\Delta t}{4D}\sum_{k=1}^{n}\left(\frac{q_k - q_{k-1}}{\Delta t}\right)^2 + < \dot{\bar{q}}_{k-1} >^2 \\ -\frac{\Delta t}{4D}\sum_{k=1}^{n} D^{1/2}\Delta \mathcal{W}_k < \ddot{\bar{q}}_{k-1} >_{(q_{k-1})} \end{array}\right] \quad (64)$$

$$= \int_{q_0}^{q} \mathcal{D}q\, exp\left[-\frac{1}{2D}\sum_{k=1}^{n} < \dot{\bar{q}}_{k-1} > \Delta q_k\right] exp - \frac{\Delta t}{4D}\left[\sum_{k=1}^{n}\begin{array}{l}\left(\frac{q_k - q_{k-1}}{\Delta t}\right)^2 + < \dot{\bar{q}}_{k-1} >^2 \\ -2D\frac{\partial < \dot{\bar{q}}_{k-1} >}{\partial \bar{q}_{k-1}}\end{array}\right]$$

where, for the mid-point approximation we have that $\bar{q}_k = \frac{q_{k+1} + q_k}{2}$, $< \dot{\bar{q}}_k > = < \frac{\dot{q}_{k+1} + \dot{q}_k}{2} >$ and $< \ddot{\bar{q}}_k > = < \frac{\ddot{q}_{+1} + \ddot{q}_k}{2} >$.

Moreover, in the continuous limit (i.e., $lim_{\Delta t\to 0}\left\{\Pi_{k=1}^{n}\int_{-\infty}^{\infty} dq_{k-1}\right\}\left(4\text{®} D\Delta t\right)^{-n/2} = \int_{q_0}^{q}\mathcal{D}q$, $lim_{\Delta t\to 0}\left(q_{k-1} - < q_{k-1} >\right) = 0$ and $lim_{\Delta t\to 0}\left(< \dot{q}_{k-1} > - < \dot{\bar{q}}_{k-1} >\right) = 0$) (64) leads to

$$P(q,q_0 \mid t-t_0,t_0) = \left(exp\int_{q_0}^{q}\frac{1}{2D} < \dot{q} >_{(q,t)} dq\right)\int_{q_0}^{q}\mathcal{D}q\, exp - \frac{1}{4D}\int_{t_0}^{t} dt\left(\dot{q}^2 + < \dot{q} > +2D\frac{\partial < \dot{q} >}{\partial q}\right) \quad (65)$$

Moreover, by comparing (64) with (D.8) , the discrete PTF $\mathcal{P}(q_k, q_{k-1} \mid \Delta t, (k-1)\Delta t)$, defined in Appendix D reads

$$lim_{\Delta t \to 0} \mathcal{P}( q_k ,q_{k-1} | \Delta t,( k-1)\Delta t ) = \left(4® D\Delta t\right)^{-n/2} exp \frac{\Delta t}{2D} \left[ \begin{array}{l} - < \dot{\bar{q}}_{k-1} > \left( \frac{q_k - q_{k-1}}{\Delta t} \right) \\ -\frac{1}{2} \left( \begin{array}{l} \left( \frac{q_k - q_{k-1}}{\Delta t} \right)^2 + < \dot{\bar{q}}_{k-1} >^2 \\ -2D \frac{\partial < \dot{\bar{q}}_{k-1} >}{\partial \bar{q}_{k-1}} \end{array} \right) \end{array} \right]$$

$$= \left(4® D\Delta t\right)^{-n/2} exp \frac{\Delta t}{4D} \left[ -\left( \dot{q}_{k-1} - < \dot{\bar{q}}_{k-1} > \right)^2 + 2D \frac{\partial < \dot{\bar{q}}_{k-1} >}{\partial \bar{q}_{k-1}} \right]$$

(66)

By analyzing (66) we observe that, due to the quantum potential $V_{qu( n_{q_k} )}$ , $< \dot{\bar{q}}_{k-1} >$ and $\frac{\partial < \dot{\bar{q}}_{k-1} >}{\partial q_{k-1}}$ depend by $n( q_k ,k\Delta t )$ and $n( q_{k+1} ,(k+1)\Delta t )$ that are unknown at the time $(k-1)\Delta t$ .

It is possible to overcome this problem if, the discrete time interval $\Delta t$ is sufficiently small to satisfying the condition (63) so that we can procede by successive steps of approximation, by using at zero order, the PTF

$$\mathcal{P}^{(0)}( q_k ,q_{k-1} | \Delta t,( k-1)\Delta t )$$
$$= \left(4® D\Delta t\right)^{-1/2} exp \frac{\Delta t}{4D} \left[ -\left( \dot{q}_{k-1} - \frac{< \dot{q}_k > + < \dot{q}_{k-1} >}{2} \right)^2 + D \frac{\partial < \dot{q}_k > + < \dot{q}_{k-1} >}{\partial \bar{q}_{k-1}} \right] \quad (67)$$
$$\cong \left(4® D\Delta t\right)^{-1/2} exp \frac{\Delta t}{4D} \left[ -\left( \dot{q}_{k-1} - < \dot{q}_{k-1} > \right)^2 + 2D \frac{\partial < \dot{q}_{k-1} >}{\partial q_{k-1}} \right]$$

that can be used to find the zero order of approximation of the PMD at subsequent time instant

$$n^{(0)}( q_k ,k\Delta t ) = \int_{-\infty}^{\infty} \mathcal{P}^{(0)}( q_k ,q_{k-1} | \Delta t,\left(k-1\right)\Delta t ) n( q_{k-1} ,\left(k-1\right)\Delta t ) dq_{k-1} \quad (68)$$

that allows to find an approximated values of the quantum potential at the $k$-instant and to calculate

$$< \dot{q}_k >^{(0)} = -\frac{1}{m©} \frac{\partial \left( V_{(q_k)} - (\frac{\hbar^2}{4m}) \left( \left( \nabla^2 ln\, n^{(0)}{}_{(q_k)} \right) - \frac{\left( \nabla ln\, n^{(0)}{}_{(q_k)} \right)^2}{2} \right) \right)}{\partial q_k} . \quad (69)$$

Thence, in the small noise limit (i.e., $D \to 0$ ) satisfying conditions (25, 63,), the output (69) allows us to obtain, at the succesive order of approximation, a more accurate PTF that reads

$$\mathcal{P}^{(0.1)}(q_k, q_{k-1} \mid \Delta t, (k-1)\Delta t)$$
$$= \left(4\text{®} D\Delta t\right)^{-1/2} exp \frac{\Delta t}{4D}\left[ -\left( \dot{q}_{k-1} - \frac{< \dot{q}_k >^{(0)} + < \dot{q}_{k-1} >}{2} \right)^2 + D \frac{\partial < \dot{q}_k >^{(0)} + < \dot{q}_{k-1} >}{\partial \overline{q}_{k-1}} \right]$$
$$\cong \left(4\text{®} D\Delta t\right)^{-1/2} exp \frac{\Delta t}{4D}\left[ -\left( \dot{q}_{k-1} - \frac{< \dot{q}_k >^{(0)} + < \dot{q}_{k-1} >}{2} \right)^2 + 2D \left( \frac{\partial < \dot{q}_k >^{(0)}}{\partial q_k} + \frac{\partial < \dot{q}_{k-1} >}{\partial q_{k-1}} \right) \right]$$

(70)

as well as the associated PDF

$$n^{(0.1)}(q_k, k\Delta t) = \int_{-\infty}^{\infty} \mathcal{P}^{(0.)}(q_k, q_{k-1} \mid \Delta t, (k-1)\Delta t) n(q_{k-1}, (k-1)\Delta t) dq_k \qquad (71)$$

and the mean velocity

$$< \dot{q}_k >^{(0.1)} = -\frac{1}{m\text{©}} \frac{\partial \left( V_{(q_k)} - (\frac{\hbar^2}{4m}) \left( \left( \frac{\partial^2 ln\, n^{(0.1)}{}_{(q_k)}}{\partial q^2} \right) - \frac{1}{2} \left( \frac{\partial}{\partial q} ln\, n^{(0.1)}{}_{(q_k)} \right)^2 \right) \right)}{\partial q_k}. \qquad (72)$$

Thence, by calculating the PTF and the related physical quantities at successive $u$-th order of approximation ($u$=2,,3,........$r$)

$$\mathcal{P}^{(0.u)}(q_k, q_{k-1} \mid \Delta t, (k-1)\Delta t)$$
$$= \left(4\text{®} D\Delta t\right)^{-1/2} exp \frac{\Delta t}{4D}\left[ -\left( \dot{q}_{k-1} - \frac{< \dot{q}_k >^{(0.(u-1))} + < \dot{q}_{k-1} >}{2} \right)^2 + D \frac{\partial < \dot{q}_k >^{(0.(u-1))} + < \dot{q}_{k-1} >}{\partial \overline{q}_{k-1}} \right]$$
$$\cong \left(4\text{®} D\Delta t\right)^{-1/2} exp \frac{\Delta t}{4D}\left[ -\left( \dot{q}_{k-1} - \frac{< \dot{q}_k >^{(0.(u-1))} + < \dot{q}_{k-1} >}{2} \right)^2 + D \left( \frac{\partial < \dot{q}_k >^{(0.(u-1))}}{\partial q_k} + \frac{\partial < \dot{q}_{k-1} >}{\partial q_{k-1}} \right) \right]$$

(73)

$$n^{(0.u)}(q_k, k\Delta t) = \int_{-\infty}^{\infty} \mathcal{P}^{(0.u)}(q_k, q_{k-1} \mid \Delta t, (k-1)\Delta t) n(q_{k-1}, (k-1)\Delta t) dq_k \qquad (74)$$

$$< \dot{q}_k >^{(0.u)} = -\frac{1}{m\text{©}} \frac{\partial \left( V_{(q_k)} - (\frac{\hbar^2}{4m}) \left( \left( \frac{\partial^2 ln\, n^{(0.u)}{}_{(q_k)}}{\partial q^2} \right) - \frac{1}{2} \left( \frac{\partial}{\partial q} ln\, n^{(0.u)}{}_{(q_k)} \right)^2 \right) \right)}{\partial q_k}, \qquad (75)$$

by iteration, we can derive the fully PTF $\mathcal{P}(q_k, q_{k-1} | \Delta t, (k-1)\Delta t)$ that reads

$$\mathcal{P}(q_k, q_{k-1} | \Delta t, (k-1)\Delta t) \cong \mathcal{P}^{(r+1)}(q_k, q_{k-1} | \Delta t, (k-1)\Delta t)$$

$$\cong \left(4\text{®} D\Delta t\right)^{-1/2} exp \frac{\Delta t}{4D}\left[ -\left( \dot{q}_{k-1} - \frac{<\dot{q}_k>^{(0.r)} + <\dot{q}_{k-1}>}{2} \right)^2 + D\left( \frac{\partial <\dot{q}_k>^{(0.r)}}{\partial q_k} + \frac{\partial <\dot{q}_{k-1}>}{\partial q_{k-1}} \right) \right] \cdot \qquad (76)$$

Where the value of *r* is defined when $<\dot{q}_k>^{(0.(r+1))} - <\dot{q}_k>^{(0.r)}$ becomes sufficiently small.

Finally, by using $\mathcal{P}(q_k, q_{k-1} | \Delta t, (k-1)\Delta t)$, the PDF at he $k$ -time instant reads

$$n(q_k, k\Delta t) = \int_{-\infty}^{\infty} \mathcal{P}(q_k, q_{k-1} | \Delta t, (k-1)\Delta t) n(q_{k-1}, (k-1)\Delta t) dq_{k-1} \cdot \qquad (77)$$

The PDFs at the successive time $k+1$ are obtained by repeating the procedure from (68).

## 3.3. The role of the initial condition in the evolution of superposition of states

Given the procedure for calculating the evolution of the PMD $n_{(q,t)}$ from the initial distribution $n_{(q,t_0)}$ we investigate in this section how the final stationary configuration depends by the initial time $t_0$.

A simple example is considered by taking the generic superposition of three eigenstates of a harmonic oscillator

$$|\cdot> = \frac{a}{\sqrt{|a|^2 + |b|^2 + |c|^2}} |\cdot_0> + \frac{b}{\sqrt{|a|^2 + |b|^2 + |c|^2}} |\cdot_1> + \frac{c}{\sqrt{|a|^2 + |b|^2 + |c|^2}} |\cdot_2> \qquad (78)$$

where $a, b, c \in \mathbb{C}$, defines the PMD and the velocity field at the initial time..

Without loss of generality, we can assume that:

i. for $t < t_0$ the quantum oscillator udergoes the deterministic quantum mechanical evolution (we pose at initial time: $a = |a| e^{i\frac{E_1{}^2 1}{\hbar}}$, $b = |b| e^{i\frac{E_2{}^2 2}{\hbar}}$, $c = |c| e^{i\frac{E_3{}^2 3}{\hbar}}$ )
ii. at the initial time $t_0$ the system is put in contact with the environmental fluctuations
iii. for $t > t_0$ it undergoes the quantum-stochastic relaxation.

As a function of the initial time $t_0$, the probability mass density PMD reads

$$n_{(q,t_0)} = |\cdot\ |^2 = \left(\frac{m\mu}{\hbar ®}\right)^{1/2} \frac{exp[-\frac{m\mu}{h}q^2\,]}{|a|^2+|b|^2+|c|^2} \left( \left\{ \begin{array}{l} +|a|\,cos[-\frac{\mu}{2}t_0+\mu{}^{2}{}_{1}\,] \\ +\frac{|b|}{\sqrt{2}}q\,cos[-\frac{3\mu}{2}t_0+3\mu{}^{2}{}_{2}\,] \\ +\frac{|c|}{2\sqrt{2}}\left(4q^2-2\right)cos[-\frac{5\mu}{2}t_0+5\mu{}^{2}{}_{3}\,] \end{array} \right\}^2 + \left\{ \begin{array}{l} |a|\,sin[-\frac{\mu}{2}t_0+\mu{}^{2}{}_{1}\,] \\ +\frac{|b|}{\sqrt{2}}q\,sin[-\frac{3\mu}{2}t_0+3\mu{}^{2}{}_{2}\,] \\ +\frac{|c|}{2\sqrt{2}}\left(4q^2-2\right)sin[-\frac{5\mu}{2}t_0+5\mu{}^{2}{}_{3}\,] \end{array} \right\}^2 \right) \quad (79)$$

that, being $\mathcal{L}^2 \approx 2\Delta q_0^2 = \frac{2\hbar}{m\mu}$, leads to

$$© = \frac{8\hbar\sim}{m\mathcal{L}^2} = 4\sim\mu \quad (80)$$

$$D = \frac{kT}{2m\mu} = \frac{\hbar}{2m}\frac{kT}{\hbar\mu} \quad , \quad (81)$$

Moreover, by using the identity

$$ln\, n_{(q,t_0)} = ln\, n_{1(q)} + ln\, n_{2(q,t_0)}\,, \quad (82)$$

where

$$ln\, n_{1(q)} = \frac{1}{2}ln\left(\frac{m\mu}{\hbar ®}\right) - \frac{m\mu}{h}q^2 \quad (83)$$

and

$$ln\, n_{2(q,t_0)} = ln\left[ \frac{1}{|a|^2+|b|^2+|c|^2} \left( \begin{array}{l} |a|^2 + \frac{|b|^2}{2}q^2 + \frac{|c|^2\left(4q^2-2\right)^2}{8} \\ +2|a|\,cos[-\mu\left(\frac{t_0}{2}+{}^{2}{}_{1}\right)]\,\frac{|b|}{\sqrt{2}}q\,cos[-3\mu\left(\frac{t_0}{2}+{}^{2}{}_{2}\right)] \\ +2|a|\,cos[-\mu\left(\frac{t_0}{2}+{}^{2}{}_{1}\right)]\,\frac{|c|}{2\sqrt{2}}\left(4q^2-2\right)cos[-5\mu\left(\frac{t_0}{2}+{}^{2}{}_{3}\right)] \\ +\frac{|b|}{\sqrt{2}}q\,cos[-3\mu\left(\frac{t_0}{2}+{}^{2}{}_{2}\right)]\,\frac{|c|}{2\sqrt{2}}\left(4q^2-2\right)cos[-5\mu\left(\frac{t_0}{2}+{}^{2}{}_{3}\right)] \\ +2|a|\,sin[-\mu\left(\frac{t_0}{2}+{}^{2}{}_{1}\right)]\,\frac{|b|}{\sqrt{2}}q\,sin[-3\mu\left(\frac{t_0}{2}+{}^{2}{}_{2}\right)] \\ +2|a|\,sin[-\mu\left(\frac{t_0}{2}+{}^{2}{}_{1}\right)]\,\frac{|c|}{2\sqrt{2}}\left(4q^2-2\right)sin[-5\mu\left(\frac{t_0}{2}+{}^{2}{}_{3}\right)] \\ +\frac{|b|}{\sqrt{2}}q\,sin[-3\mu\left(\frac{t_0}{2}+{}^{2}{}_{2}\right)]\,\frac{|c|}{2\sqrt{2}}\left(4q^2-2\right)sin[-5\mu\left(\frac{t_0}{2}+{}^{2}{}_{3}\right)] \end{array} \right) \right]$$

(84)

that, being from (83)

$$\frac{1}{m©}\frac{\partial\left(V_{(q)}-(\frac{\hbar^2}{4m})\left(\left(\frac{\partial^2 ln\, n_{1(q)}}{\partial q^2}\right)-\frac{1}{2}\left(\frac{\partial}{\partial q}ln\, n_{1(q)}\right)^2\right)\right)}{\partial q}=0' \tag{85}$$

leads to the velocity field at initial time

$$<\dot{q}_{(q,t_0)}>=\frac{1}{©}\frac{\hbar^2}{4m^2}\frac{\partial\left(\left(\frac{\partial^2 ln\, n_{2(q,t_0)}}{\partial q^2}\right)-\frac{1}{2}\left(\frac{\partial}{\partial q}ln\, n_{2(q,t_0)}\right)^2\right)}{\partial q} \tag{86}$$

that depends by the initial time $t_0$.

### 3.4. General characteristics of the quantum relaxation

In absence of the quantum potential, the Brownian process described by the FPE admits the stationary long-time solution

$$P(q,q_{-\infty}|t-t_{-\infty},t_{-\infty})=N\,exp\frac{1}{D}\int_{q=-\infty}^{q}<\dot{q}>_{(q,t)t_0\to-\infty}dq'=N\,exp\frac{1}{D}\int_{q=-\infty}^{q}K(q')dq' \tag{87}$$

leading to the canonical expression [47]

$$P(q,q_0|t-t_0,t_0)=\left(exp\int_{q_0}^{q}\frac{1}{2D}K(q')dq'\right)\int_{q_0}^{q}\mathcal{D}q\,exp-\frac{1}{4D}\int_{t_0}^{t}dt\left(\dot{q}^2+K(q)+2D\frac{\partial K(q)}{\partial q}\right) \tag{88}$$

Generally speaking, in the quantum case, $<\dot{q}>_{(q,t)}$ owns a functional dependence by $n_{(q,t)}$ and it is not possible to obtain the general form of the term $\frac{1}{2D}<\dot{q}>_{(q,t)}$ since it depends by the specific relaxation path of the system, $n_{(q,t,t_0)}$, toward the stationary state that in non-stationary quantum superposition of states sensibly depends by $t_0$.

Thence, in principle, different long-time stationary configurations $n_{(q,t=\infty)}$ could be reached, as a function of the initial time $t_0$, if more than one eigenstate exists.

Moreover, since the MDDs $\tilde{n}_{(q,t)}$ (realized as functions of the exact sequence of random force inputs) is connected to the PMD only at first approximation of $\dot{q}$, in the limit of slow relaxation process, and small fluctuations amplitude [42], in the case of a large fluctuation (that can happen on large time scale), $\tilde{n}_{(q,t)}$ might make a transition from a stationary quasi-eigenstate to a generic superposition of states leading, through a new relaxation, to adifferent stationary quasi-eigenstate. Since during the jump process due to a large fluctuation, the PMD $n_{(q,t)}$ loses its connections with the MDD $\tilde{n}_{(q,t)}$, the PMD can only catch the relaxation process after each large fluctuation, but not the complete system evolution.

In a collection of a large number of particles where, on short time scale, each one relaxes to an eigenstate, while on long time scale they can randomly make transitions to other eigenstates, we have the generation of the re-shuffling of the state of all particles leading to the establishment of the statistical distribution (that canno be described by the PMD).

## 4. Discussion

Before analyzing the features of the SQHM it is worthmentioning to observe that equation (1-3) constitute an approximated version of the model. Aside from the approximation (4), that leads to the Langevin Schrodinger equation (and, under certain conditions [42], to coarse grained macroscopic classical mechanics), the constancy of © is a quite subtle assumption given that, generally speaking, it is not constant [42, 44-45]. The goal of this guesswork is mainly due to obtain a manageable model able to capture the peculiarities of the quantum relaxation in presence of noise.
Another not well seizable parameter, introduced into the theory, is the dissipation parameter ~ that, in principle, should be self-defined by the theory but that, here, is semiempirically presumed. The assumption of its smallness for microscopic systems close to the quantum "deterministic" limit is of common sense (since condition (18) has its foundation from the empirical information that wants the quantum mechanics realized for $a_c \to \infty$ or $T \to 0$) but the physical order of magnitude is not known and has to be measured.
In principle ~ is not a constant and depends both by the form of the Hamiltonian interaction (determining the Lyapunov exponents) and, in chaotic systems, by the noise amplitude.
Finally, it must be noted that the assumption that PMD $n$ evolution is physically meaningful and contains information about the MDD $\tilde{n}$, in the limit of small noise amplitude, is warranted by the condition $lim_{T\to 0}\, n = lim_{T\to 0}\, \tilde{n} = |\cdot|^2$ but, genrally speaking, the basin of convergence is not well defined.

### 4.1 Compatibility with the Copenhagen interpretation of QM

Firstly, it is noteworthy to observe that the Madelung approach is different from the Bhom theory [32] and it does not contain any hidden variables being perfectly equivalent to the Schrodinger model once the quantization conditions are applied (see appendix B).
Secondly, we must be aware that the path-integral solution is not general but holds in the small noise limit and for an appropriate period of time (before a large fluctuation happens). After that, in the following transient relaxation process, the SQHM shows that, in principle, the final stationary state depends by the initial condition at the time $t_0$ (e.g., (75)) allowing the system possibly to reach eachone of the eigenstates of the superposition of states.
Since the quantum superposition of state owns a cyclic evolution with period

$$\mathrm{T} = \frac{2®\hbar}{E_{min}} \qquad (89)$$

where $E_{min}$ is its lowest energy eigenvalue of the state superposition, it follows that the SQHM model fulfils the Copenhgen interpretation of quantum mechanics if, given $N$ stochastic values of the initial time $t_0$ such as $0 \le t_0 < \mathrm{T}$, the number $\frac{N_i}{N}$ (where $N_i$ is the number of times when the final stationary state is the $i$-th eigenstate) approaches to the Born rule

$$lim_{N\to\infty} \frac{N_i}{N} = \frac{|a_i|^2}{\sum_{k=1}^{t} |a_k|^2} . \qquad (90)$$

## 4.2 Wave-particle duality breaking in the SQHM

By using the continuos limit of (??), we can see how the wave-paticle equivalence is achieved in the deterministic limit given by the Madelung quantum hydrodynamics. In fact, by considering

$$
\begin{aligned}
& lim_{D\to 0}\, lim_{\Delta t\to 0}\, \mathcal{P}( q_k , q_{k-1} \,|\, \Delta t, (k-1)\Delta t ) \\
&= lim_{D\to 0}\, lim_{\Delta t\to 0} \left(4 ® D\Delta t\right)^{-n/2} exp \frac{\Delta t}{4D}\left[ -\left(\dot{q} - <\dot{q}>\right)^2 + 2D\frac{\partial <\dot{q}>}{\partial q}\right] \\
&= lim_{D\to 0}\, lim_{\Delta t\to 0} \left(4 ® D\Delta t^{-1/2}\right)^{-1/2} exp \frac{\Delta t}{4D}\left[ -\left(\dot{q} - <\dot{q}>\right)^2 \right] \\
&= -lim_{\Delta t\to 0}\, ¢ \left(\Delta t^2 \left(\dot{q} - <\dot{q}>_{D=0}\right)^2\right) exp \frac{1}{2}\left[\Delta t \frac{\partial <\dot{q}>}{\partial q}\right] \qquad , \qquad (91) \\
&= lim_{\Delta t\to 0}\, ¢ \left(\Delta t^2 \left(\frac{p}{m} - <\dot{q}>_{D=0}\right)^2\right) \\
&= lim_{\Delta t\to 0}\, ¢ \left(\Delta t^2 \left(\frac{1}{m}\frac{\partial S}{\partial q} - <\dot{q}>_{D=0}\right)^2\right)
\end{aligned}
$$

that for $\Delta t \neq 0$ leads to

$$lim_{D\to 0}\, m <\dot{q}_i> = m\dot{q}_i = \frac{\partial S}{\partial q_i} = p_i , \qquad (92)$$

to

$$lim_{D\to 0}\, n = n_{det} \qquad (93)$$

and to

$$lim_{D\to 0}\, © m = 0 \qquad (94)$$

and, besides, given that

$$lim_{D\to 0}\, P((q,p),z \,|\, t,0) = P((q, ¢\left(\frac{\partial S}{\partial q} - p\right)), z \,|\, t,0) , \qquad (95)$$

and that

$$
\begin{aligned}
lim_{D\to 0}\, n_{(q,t)} = n_{det(q,t)} &= lim_{D_p\to 0} \int N(q,p,t) d^{3h} p \\
&= \int lim_{D\to 0} \int P((q, ¢\left(\frac{\partial S}{\partial q} - p\right)), z \,|\, t,0) N_{(z,0)} d^{6h} z d^{3h} p \qquad (96) \\
&= \int N(q, ¢\left(\frac{\partial S}{\partial q} - p\right), t) d^{3h} p
\end{aligned}
$$

we obtain that, if at initial time $t_0$ the MDD $N$ of the system owns the wave-particle equivalence

$$N(q,¢\left(\frac{\partial S}{\partial q}-p\right),t_0)=n_{det(q,t_0)}¢\left(\frac{\partial S}{\partial q}-p\right), \tag{97}$$

then, by (94) the form

$$N(q,¢\left(\frac{\partial S}{\partial q}-p\right),t)=n_{det(q,t)}¢\left(\frac{\partial S}{\partial q}-p\right). \tag{98}$$

determining the wave particle equivalence, is maintained along time.

On the other hand, generally speaking, given that the noise generates an enlargement of the ¢ -peaked $N(q,p,t)$ phase space distribution (98) centered into the subspace $\frac{\partial S}{\partial q}=p$, the wave-particle equivalence is not maintained along time.

In presence of noise, the wave momentum $\frac{\partial S}{\partial q_i}$ differs by the local momentum of the particle $m<\dot{q}_i>$,

## 4.3 The emerging of classical behaviour on coarse-grained large scale

Is matter of fact that, if the quantum potential is cancelled by hand in the quantum hydrodynamic equations of motion (1-3), the classical equation of motion emerges [30]. Even if this is true, this operation is not mathematically correct since it changes the characteristics of the QHA equations. Doing so, the stationary configurations (i.e., eigenstates) are wiped out because we cancel the balancement of the MDD against the hamiltonian forces. An even small quatum potential cannot be neglected into the QHA model.

Moreover, even if the noise $´_{(q,t,T)}$ has zero mean, the mean of the quantum potential fluctuations, $\overline{V}_{st(n,S)}\cong S$, is not zero, so that the noise effect on $<\dot{q}>$ and $<\ddot{q}>$ is not equal to that one generated the quantum potential alone as in the canonical (deterministic) quantum mechanics.

The stochastic sequence of inputs of noise alters the coherent reconstruction of superposition of state. In the SQHM this effect is generated by the dissipative force $-©\dot{q}_{(t)}$ in (2). Moreover, when the non-local force generated by the quantum potential is quite small (respect to the fluctuations amplitude) so that

$$lim_{q\to\Delta q}\left|\frac{1}{m}\frac{\partial V_{qu(n)}}{\partial q_i}\right|\ll ©\left(\frac{£}{ª_c}\right)\left(\frac{\hbar}{2m}\right)^{1/2}=©\frac{£}{2}\sqrt{\frac{kT}{2\hbar}}, \tag{99}$$

its effect can be disregarded in the coarse grained approximation of a large sistem whose resolution size $\Delta q$ (e.g., the side of the minimum cube of volume) is larger than the quantum potential range of interaction [42],. leading to the stochastic classical mechanics. As a consequence of (99), at glance with the outputs of the current literature, linear systems do not admit the large scale coarse grained classical limit being $\frac{\partial V_{qu(n)}}{\partial q_i}\propto q$ [42].

## 4.5 Measurement process and quantum decoherence

The SQHM shows that a decoherent process by itself is not necessarily a measure- Nevertheless, decoherence is necessary to pass from the quantum superposition of states to the statistical melange: The sensing part of the mesuring apparatus (pointer) and the measured system may have a canonical quantum interaction that after the measurement process, when the mesuring apparatus is brought to the infinity (at a distance much beyond $a_c$ ), ends. Then the reading and the treatment of the "pointer" state is done by the measurement apparatus: This process is an irreversible process (with a defined arrow of time) leading to the macroscopic output of the measure.
On the other hand, the breaking of the coherent quantum potential interaction (on large scale) is necessary for the measurement process in order to have, both before the initial time and after the final one, the quantum-decoupling between the measurement apparatus and the system that will allow to collect a statistical ensemble of data from repeated experiments.

### 4.6 Minimum measurements uncertainty in a quantum-stochastic super-system

If on distances smaller than $a_c$ any system is quantum so that its subparts are not independent eachother, it follows that in order to perform the measurement (with independence between the measuring apparatus and the measured system) it is necessary that they must be far apart (at least) more than $a_c$ and hence, for the finite speed of propagation of interactions and information, the measure process must last longer than the time

$$\Delta 2 = \frac{a_c}{c} = \frac{2\hbar}{(2mc^2kT)^{1/2}} \quad . \tag{100}$$

Moreover, since higher is the amplitude of the noise $T$, lower is the value of $a_c$ and higher are the fluctuations of the energy measurements $\Delta E_{(T)}$, it follows that the minimum duration of the measurement $\Delta 2 = \frac{a_c}{c}$ multiplied by the precision of the energy measurement $\Delta E_{(T)}$ has a lower bond.

Given the Gaussian noise (9,20), we have that the mean value of the energy fluctuation is $\Delta E_{(T)} = kT$ . Thence, for the non-relativistic case ( $mc^2 >> kT$ ) a particle of mass $m$ owns an energy variance $\Delta E$

$$\begin{aligned}\Delta E &\approx (<(mc^2+\Delta E_{(T)})^2-(mc^2)^2>)^{1/2} \cong (<(mc^2)^2+2\Delta E-(mc^2)^2>)^{1/2} \\ &\cong (2mc^2<\Delta E>)^{1/2} \cong (2mc^2kT)^{1/2}\end{aligned} \tag{101}$$

from which we can evaluate that

$$\Delta E\Delta t > \Delta E\Delta 2 \approx \frac{(2mc^2kT)^{1/2} a_c}{c} \approx \frac{2}{®} h \, , \tag{102}$$

It is worth noting that the product $\Delta E\Delta 2$ is constant since the growing of the energy variance with the square root of $T$ is exactly compensated the decrease of the minimum time 2 of measurement

The same result is achieved if we derive the uncertainty relations between the position and momentum of a particle of mass $m$. If we measure the spatial position of a particle with a precision $\Delta L > a_c$ so that we do not perturb the measured quantum system, the variance $\Delta p$ of the modulus of its relativistic momentum $( p^{«} p_{«} )^{1/2} = mc$ due to the fluctuations reads0

$$\begin{aligned}\Delta p &\approx ( < ( mc + \frac{\Delta E_{(T)}}{c} )^2 - ( mc )^2 > )^{1/2} \cong ( < ( mc )^2 + 2m\Delta E - ( mc )^2 > )^{1/2} \\ &\cong ( 2m < \Delta E > )^{1/2} \cong ( 2mkT )^{1/2}\end{aligned} \quad (103)$$

leading to the uncertainty relation

$$\Delta L \Delta p > a_c ( 2mkT )^{1/2} \approx \frac{2}{®} h \quad (104)$$

If we want to measure the spatial position with a precision $\Delta L < a_c$, we have to perturb the quantum state of the particle. Due to the increase of the spatial confinement of the wave function, an increase of both the quantum potential energy and its fluctuations are generated so that the final particle momentum gets a variance $\Delta p$ higher than (103) but submitted to the theoretical quantum uncertainty constraint.

As far as it concerns the *theoretical minimum uncertainty* of quantum mechanics, obtainable from the *minimum measurements uncertainty* (102,104) in the limit of zero noise, we observe that even if the quantum deterministic behavior (for $a_c \to \infty$) in the low velocity limit (i.e., $c \to \infty$) leads to the undetermined inequalities

$$\Delta^2 \geq \frac{a_c}{c} \quad (105)$$

$$\Delta E \cong ( mc^2 kT )^{1/2} = \sqrt{2}\hbar \frac{c}{a_c} \quad (106)$$

it owns the product

$$\Delta^2 \Delta E \geq \sqrt{2}\hbar \quad (107)$$

defined, disemboguing into the *minimum uncertainty* of the quantum mechanics.

Since, as (105) shows, the duration of the measurement in the deterministic limit becomes infinite and the process is endless and cannot be performed, stricly speaking, (107) cannot be referred as the *minimum measurements uncertainty* but the *minimum uncertainty* of the quantum state.

Moreover, since both non-locality is confined in domains of physical lenght smaller $a_c$ (or smaller than the quantm potential range of interaction [42]) and information about a quantum system cannot be transferred faster than the light speed (otherwise also the uncertainty principle will be violated) the local realism is obtained in the coarse-grained large scale physics and the paradox of the "spooky action at a distance " is limited on microscopic distance smaller than $a_c$ (or than the quantum potential range of interaction [42]).

## 4.7 Fields of Application of the SQHM

The theory that describes how the quantum entanglement extends itself up to a cetrain distance and how to minimizing the quantum decoherence, can lead to important improvements in the development of materials for high-temperature superconductors, Qbits systems and in the description of kinetics of complex chemical reactions at the edge of the quantum to classical regime. Generally speaking, the SQHM can furnish an analytical self-consistent theoretical model for mesoscale phenomena and quantum irreversibility.

## 5. Conclusions

By starting from the stochastic model of the Madelung quantum hydrodynamics the evolution of the quantum sistems is analyzed.
When the quantum eigenstates are submitted to small fluctuations their stationary configurations are sligthly perturbed and remain close to that ones of quantum mechanics. On the contrary, the evolution of the superposition of states is deeply perturbed showing that they can relax to the stationary configuration reconducible to one of the quantum eigenstates.
The stochastic quantum hydrodynamic model allows to identify the eigenstates from their intrinsic properties of stationarity and stability under small fluctuations avoiding their definition via the measurement process (making the theory independent by classical operations).
The SQHM shows that for defining the relaxation pathway of the entangled quantum configuration, the state at the initial time $t_0$ (when the interaction either with the classical environment or the experimental apparatus for the measurement starts) is needed.

The SQHM unifies the hydrodynamic description of quantum mechanics with the decoherence approach showing that the agreement with the Copenhagen interpretation of quantum mechancis is also possible.
The SQHM shows that the minimum uncertainty in the measurement process is warranted if interactions and information do not travel faster than the speed.of light, showing that the large scale locality and the relativistic postulate are compatible with the non-local quantum interactions at the microscale: The minimum measurement uncertainty leads to the theoretical minimum uncertainty of quantum mechanics in the limit of zero noise.

# Appendix A

## Quantum (immaginary-time) stochastic process

The Schrödinger equation

$$i\hbar\frac{\partial \cdot}{\partial t} = -\frac{\hbar^2}{2m}\frac{\partial^2 \cdot}{\partial q_i \partial q_i} + U_{(q)} \cdot \qquad \text{(A.1)}$$

in the quantum integral path representation, is equivalent to the FPE [1 klinert] that in the unidimentional case reads

$$\frac{\partial P_{(q,t|\, q_a,t_a)}}{\partial t} = -\frac{\partial K(q) P_{(q,t|\, q_a,t_a)}}{\partial q} + D\frac{\partial^2 P_{(q,t|\, q_a,t_a)}}{\partial q^2} \qquad \text{(A.2)}$$

where the solution

$$P_{(q,t|\, q_a,t_a)} = \sqrt{\frac{P_{s(q)}}{P_{s(q_a)}}} \int_{q_a}^{v} Dq \quad exp\left[-\frac{1}{2D}\int_{t_a}^{t} dt\left(\frac{1}{2}\dot{q}^2_{(t)} + \frac{1}{2}K_{(q)} + DK'_{(q)}\right)\right], \text{(A.3)}$$

where

$$P_{s(x)} = \qquad \text{(A.4)}$$

and

$$K'_{(q)} = \frac{\partial K_{(q)}}{\partial q}, \qquad \text{(A.5)}$$

leads to the imaginary time evolution amplitude (i.e., the wave function)

$$\cdot_{(q,t)} = \cdot_{(q,0)} \int Dq \quad exp\left[-\frac{1}{\hbar}\int_0^{-i^2} d^2 \left(\frac{M}{2}\dot{q}^2_{(^2)} + V_{(q(^2))}\right)\right] = \sqrt{\frac{P_{s(q_a)}}{P_{s(q)}}} P_{(q,t|\, q_a,t_a)} \text{(A.6)}$$

Equation (A.2) refers to a Brownian particle with mass $m$ and friction coefficient $= ©m$ in an external potential $U_{(q)}$ obeying to the SDE

$$\dot{q}_j = -\frac{1}{}\frac{\partial U_{(q)}}{\partial q_j} + D^{1/2}_{jm}¶_m \qquad \text{(A.7)}$$

such as $\frac{\partial U_{(q)}}{\partial x} = -m© K_{(q)}$ where $D_{im} = D¢_{jm} = \frac{\hbar}{2m}¢_{jm}$ and $^2 = it$, that is equivalent to a quantum particle submitted to the physical potential $V_{(q)}$ that reads

$$V_{(q)} = \frac{m}{2}K^2_{(q)} + \frac{\hbar}{2}\frac{\partial K_{(q)}}{\partial q}. \qquad \text{(A.8)}$$

The conservation equation for the mass density $n=|\cdot|^2$ can be obtained, in a simple way [SME], by using the quantum hydrodynamic model that describes the wave function $\cdot_{(q,t)}=|\cdot|_{(q,t)}\, exp[\frac{i}{\hbar}S_{(q,t)}]$ as the motion of a particle density $n_{(q,t)}=|\cdot|^2_{(q,t)}$ with velocity $\dot{q}_i=\frac{1}{m}\frac{\partial S_{(q,t)}}{\partial q_i}$ ) governed by the equations [ 4]

$$\partial_t n_{(q,t)}+\frac{\partial(n_{(q,t)}\dot{q})}{\partial q_i}=0\,, \tag{A.10}$$

$$\dot{q}_i=\frac{1}{m}\frac{\partial S_{(q,t)}}{\partial q_i}=\frac{p_i}{m}\,, \tag{A.11}$$

$$p_i=-\frac{\partial(H+V_{qu})}{\partial q_i}\,, \tag{A.12}$$

where, for a non relativistic particle in an external potential, can assume

$$H=\frac{p_i p_i}{2m}+V_{(q)} \tag{A.13}$$

and where

$$S=\int dt(\frac{p_i p_i}{2m}-V_{(q)}-V_{qu}) \quad , \tag{A.14}$$

where $V_{qu}$ is the quantum pseudo-potential that reads

$$V_{qu}=-(\frac{\hbar^2}{2m})n^{-1/2}\frac{\partial}{\partial q_j}\frac{\partial}{\partial q_j}n^{1/2}=-(\frac{\hbar^2}{2m})\left(\left(\frac{\partial}{\partial q_j}\frac{\partial}{\partial q_j}ln\, n^{1/2}\right)+\left(\frac{\partial\, ln\, n^{1/2}}{\partial q_j}\right)^2\right)$$
$$=-(\frac{\hbar^2}{2m})\frac{1}{2}\left(\left(\frac{\partial}{\partial q_j}\frac{\partial}{\partial q_j}ln\, n\right)+\frac{1}{2}\left(\frac{\partial\, ln\, n}{\partial q_j}\frac{\partial\, ln\, n}{\partial q_j}\right)\right) \tag{A.15}$$

leading to the kinetic law (see Appendix B)

$$\dot{q}=\frac{1}{m}\frac{\partial S}{\partial q_i}=\frac{1}{m}\int_{t_0}^{t}dt(\frac{\partial}{\partial q_i}\frac{p_i p_i}{2m}-\frac{\partial\left(V_{(q)}-V_{qu}\right)}{\partial q_i})$$
$$=-\frac{1}{m}\int_{t_0}^{t}dt(\frac{\partial\left(V_{(q)}-V_{qu}\right)}{\partial q_i}) \tag{A.16}$$

and to the stationary equilibrium condition $\dot{q}=0$

$$-\frac{1}{m}\int_{t_0}^{t}dt(\frac{\partial\left(V_{(q)}-V_{qu}\right)}{\partial q_i})=0 \tag{A.17}$$

that ,by using (A.15), explicitly reads

$$\frac{\partial}{\partial q_i}\left( V_{(q)} + (\frac{\hbar^2}{4m})\left( \left( \frac{\partial^2}{\partial q_i \partial q_i} ln\, n^{1/2} \right) + \frac{1}{2}\left( \frac{\partial}{\partial q_i} ln\, n^{1/2} \right)^2 \right) \right) = 0 \qquad \text{(A.18)}$$

that allows to derive the expression for the quantum eigenstates (see Appendix B)).

## Thermal stochastic process

The classical damped stochastic motion equations (for a classical mass density distribution $\overline{n}_{(q,t)}$) (e.g., the single-particle distribution approximation for rarefied gas phase and Markovian fluids [50]) reads-

$$\dot{q}_j = \frac{dq_j}{dt} = \frac{p_j}{m}$$

$$\dot{p}_j = -m\copyright \dot{q}_{j(t)} - \frac{\partial V_{(q)}}{\partial q_j} + m\copyright D^{1/2} \P_{(t)} \qquad \text{(A.19)}$$

$$\ddot{q}_{j(t)} = -\copyright \dot{q}_{j(t)} - \frac{1}{m}\frac{\partial V_{(q)}}{\partial q_j} + \copyright D^{1/2} \P_{(t)}$$

, (A.20)

where $D_j^{1/2} = \sqrt{\frac{2kT}{\ }}$ (where $= \copyright m$ is the friction coefficient), that leads to the long-time limiting equation of (A.19)

$$\dot{q}_j = -\frac{1}{\ }\frac{\partial V_{(q)}}{\partial q_j} + D^{1/2} \P_{(t)}, \qquad \text{(A.21)}$$

with the "position distribution" [49] that obeys to the Fokker-Plank equation

$$\frac{\partial}{\partial t}\overline{n}_{(q,t)} + \frac{\partial}{\partial q_j}(\overline{n}_{(q,t)} v_j) = 0 \qquad \text{(A.22)}$$

where

$$v_j = -\frac{1}{\ }\frac{\partial V_{(q)}}{\partial q_j} - \frac{1}{2\tilde{n}}\frac{\partial D\overline{n}_{(q,t)}}{\partial q_j} \qquad \text{(A.23)}$$

that, in the case of constant diffusion coefficient $D$, reads

$$v_j = -\frac{1}{\ }\frac{\partial V_{(q)}}{\partial q_j} - \frac{\mathrm{D}}{2}\frac{\partial\, ln\, \overline{n}}{\partial q_j}. \qquad \text{(A.24)}$$

In presence of local thermodynamic equilibrium (i.e, $\mathrm{D} = \frac{2kT}{\ }$), under isothermal condition, relation (A.24)reads [50]

$$v_j = -\frac{D}{2}\left( \frac{\partial\left(\frac{V}{kT}\right)}{\partial q_j} + \frac{\partial\, ln\,\overline{n}}{\partial q_j} \right) \tag{A.25}$$

that in a stationary state leads to

$$v_j = -\frac{D}{2kT}\frac{\partial}{\partial q_j}\left( V_{(q)} + kT\, ln\,\overline{n} \right) = 0 \tag{A.26}$$

The above relation shows that at equilibrium the Hamiltonian force $-\frac{\partial V_{(q)}}{\partial q_j}$ is balanced by the diffusional force $-\frac{\partial kT\, ln\,\overline{n}}{\partial q_j}$ given by gradient of the chemical potential.

By comparing (SEQ) with QEQ), we can see that the quantum imaginary stochastic process (A.10-12) can be seen as a "pseudo-diffusional" process driven by the mass density distribution through the quantum potential $\propto n^{-1/2}\frac{\partial}{\partial q_j}\frac{\partial}{\partial q_j}n^{1/2}$ instead of $\propto ln\,\overline{n}$ as in the thermal stochastic process.

The analogy (but not equivalence) between the thermal and the quantum stochastic process consists in the fact that the quantum stationary eigenstates (i.e., $\dot{q} = 0$) happens when the force generated by the quantum potential exactly counterbalances, point by point, that one given by the Hamiltonian potential $V_{(q)}$.

The basic difference is that, in the quantum case, we may have more than one stationary state because many quantum potential expressions (as a function of $n^{1/2}$) exist that counter-balance the hamiltonian potential (see Appendix B).

In absence of an external field $V_{(q)}$, in both cases the equilibrium distribution is characterized by the absence of mass density gradient and the variance of a Gaussiam mass density distriution grows with time.

Even if there exists an analogy between the real-time stochastic process (A.22-23) and the immaginary-time one (A.10-12) they own important differences.

For instance, the quantum "immaginary-time" brownian process of a free Gaussian mass-distribution $n^{1/2} = n_0\, exp\left[ -\frac{q^2}{2\Delta q^2} \right]$ leads to the spreading where

$\Delta q^2 \propto t^2$,

while the thermal fluctuations the spreading leads to the spreading

$\Delta q^2 \propto t$,

Furthermore, the "real-time" classical stochastic dynamics are dissipative while the "immaginary-time" stochastic dinamics are reversible (with time-inversion symmetry) leading to the "deterministic" quantum mechanics.

If in the classical (real time) stochasticity we have an increase of the entropy during the relaxation to the stationary equilibrium state, on the other hand, in the quantum deterministic evolution there is no loss of information and there is not a relaxation process: The superposition of states, with their complex configuration, are maintained along time and never relax to a stationary equilibrium configuration given by one of the eigenstates.
When the two stochastic processes are contemporarely present (the quantum "immaginary-time" one submitted to the thermal "real-time" one), some questions arise: Does the superposition of states relax to a stationary configuration? Are these stationary configurations related to the quantum eigenstates? And if they are, to which eigenstate relaxes?

## Appendix B

## The quantum hydrodynamic description of the harmonic oscillator

The quantum hydrodynamic equations of motion corresponding to the Schrödinger equation for the complex field

$$\cdot = |\cdot| \, exp \frac{i}{\hbar} S \,, \qquad \text{(B.1)}$$

are obtained by splitting it into two equations as a function of the variables $|\cdot|$ and $\frac{\partial S}{\partial q_i}$ [48].

The hydrodynamic approach describes the evolution of the particles density $n = |\cdot|^2$ as moving with the impulse

$$p_i = \frac{\partial S}{\partial q_i} \qquad \text{(B.2)}$$

and subject to the theory-defined quantum potential $V_{qu}$ [28-30, 48] according to

$$\dot{q}_i = \frac{p_i}{m} = \frac{1}{m} \frac{\partial S_{(q,t)}}{\partial q_i} \,, \qquad \text{(B.3)}$$

$$\dot{p}_i = -\frac{\partial ( H_{cl} + V_{qu} )}{\partial q_i} \,, \qquad \text{(B.4)}$$

where

$$V_{qu} = -\frac{\hbar^2}{2m} \frac{1}{|\cdot|} \frac{\partial^2 |\cdot|}{\partial q_i \partial q_i} \,. \qquad \text{(B.5)}$$

satisfying the conservation equation

$$\frac{\partial}{\partial t} |\cdot|^2 + \frac{\partial}{\partial q_i} ( |\cdot|^2 \dot{q}_i ) = 0 \,. \qquad \text{(B.6)}$$

Actually, the hydrodynamic system of equations broaden the solutions of the Schrödinger equation since not all momenta $p_i$, of the hydrodynamic description, are solutions of the field equation [48]. This because not all momenta $p_i$ deriving from hydrodynamic equations can be expressed as gradients of the action function $S$ .
The restriction of the hydrodynamic momenta to those ones of the Schrödinger problem derives by imposing the existence of the action S by the integrability of the impulse $p_i$ that reads [48]

$$\oint p_i dl_i = -\oint \frac{\partial S}{\partial q_i} dl_i = 0 \qquad \text{(B.7)}$$

that for systems, that do not depend explicitly by time, is satisfied by the irrotational condition

$$\pounds_{ijk} \frac{\partial S}{\partial q_j} p_k = 0 \qquad \text{(B.8)}$$

Moreover, since the action $S_{(q,t)}$ is contained in the exponential argument of the wave-function, all the multiples of $2®\hbar$, with

$$S_{n\,(q,t)} = S_{(q_0,t)} + \int_{q_0}^{q} \frac{\partial S}{\partial q_i} dl_i + 2n®\hbar = S_{(q,t)} + 2n®\hbar \qquad n = 0,\ 1,\ 2,\ 3,\ \ldots \qquad \text{(B.9)}$$

are possible, so that the action results quantized.

With the "irrotational" condition the hydrodynamic approach becomes equivalent to the Schrödinger one. Being different by the Bhom theory [32], the quantum hydrodynamic approach is fully quantum [11-14] and provides outputs that completely overlap those ones of the standard quantum treatment [28].

## Implementation of quantization in the hydrodynamic quantum equations

Even if the quantization condition, as formulated by (B.7), is external to the hydrodynamic quantum equations (B.3-6), it is implicitly introduced by the quantum potential when we find the eigenstates of the quantum problem.

In the hydrodynamic description, the eigenstates are identified by their property of stationarity that is given by the "equilibrium" condition that reads

$$\dot{p} = 0 \,. \qquad \text{(B.10)}$$

This equilibrium happens when the force generated by the quantum potential exactly counterbalances that one of the external Hamiltonian potentials in a way that.

$$-\frac{\partial (V_{(q)} + V_{qu})}{\partial q_i} = 0$$

Moreover, given for eigenstates the initial condition

$$\dot{q}_{(t=0)} = \dot{q}_0 \,, \qquad \text{(B.11)}$$

by (B.10) it follows that $\dot{q} = constant = \dot{q}_0$ where, In presence of an external potential, not function of time, $\dot{q}_0$ can be choosen equal to zero, leading to

$$\dot{q} = 0 \qquad \text{(B.12)}$$

For unidimensional systems, whose Hamiltonian reads

$$H = \frac{p^2}{2m} + V_{(q)}, \qquad \text{(B.13)}$$

by using (A.14) it follows that, for the $k$-th eigenstate, it holds

$$S_k - S_{k(q,t_0)} = \int_{t_0}^{t} dt \left( \frac{m \dot{q}_i \dot{q}_i}{2} - V_{(q)} - V_{qu(k)} \right) = -(V_{(q)} + V_{qu(k)}) \int_{t_0}^{t} dt = -E_k (t - t_0) \qquad \text{(B.14)}$$

where

$$\cdot\ E_k = V_{(q)} + V_{qu(k)} \tag{B.15}$$

Moreover, by applying the stationary condition

$$\dot{p}_i = -\frac{\partial (H + V_{qu(k)})}{\partial q_i} = 0 \tag{B.16}$$

it is possible to find the eigenstates by solving the differential equation

$$V_{(q)} + V_{qu(n)} = E_n = V_{(q)} - (\frac{\hbar^2}{2m})\frac{1}{|\cdot|}\frac{\partial^2 |\cdot|}{\partial q_i \partial q_i} \tag{B.17}$$

where $E_k$ is the energy of the $k$-th eigenstate.

For instance, in the case of an unidimensional harmonic oscillator (i.e., $V_{(q)} = \frac{m\mu^2}{2} q^2$ ) reads

$$V_{qu} = -(\frac{\hbar^2}{2m})|\cdot_k|^{-1}\frac{\partial^2 |\cdot_k|}{\partial q_i \partial q_i} = E_k - \frac{m\mu^2 q^2}{2} \tag{B.18}$$

that admits solutions of type

$$|\cdot_k|_{(q,t)} = A_{k(q)}\, exp\left(-\frac{m\mu}{2\hbar}q^2\right), \tag{B.19}$$

that leads to the $k$-th eigenfunction

$$\cdot_{k(q,t)} = |\cdot_k|_{(q,t)}\, exp[\frac{i}{\hbar}S] = A_{k(q)}\, exp\left(-\frac{m\mu}{2\hbar}q^2\right) exp\left(-\frac{iE_n t}{\hbar}\right). \tag{B.20}$$

Finally, the form of $A_k$ by introducing the quantum potential of the $k$-th eigenstate reads

$$V_{qu(k)} = -\frac{m\mu^2}{2}q^2 + \left[ n\left( \frac{\frac{m\mu}{\hbar}qA_{k-1} - 2(n-1)A_{k-2}}{A_k} \right) + \frac{1}{2} \right]\hbar\mu \tag{B.21}$$

into (B.19) to obtain

$$\begin{aligned} E_k &= V_{(q)} - \frac{m\mu^2}{2}q^2 + \left[ n\left( \frac{\frac{m\mu}{\hbar}qA_{(q)k-1} - 2(k-1)A_{(q)k-2}}{A_{(q)k}} \right) + \frac{1}{2} \right]\hbar\mu \\ &= \left[ k\left( \frac{\frac{m\mu}{\hbar}qA_{(q)k-1} - 2(k-1)A_{(q)k-2}}{A_{(q)k}} \right) + \frac{1}{2} \right]\hbar\mu \end{aligned} \tag{B.22}$$

leading to the condition

$$A_k = C\left(\frac{m\mu}{\hbar} q A_{k-1} - 2(k-1) A_{k-2}\right) \tag{B.23}$$

that is satisifed by

$$A_{k(q)} = H_{k\left(\frac{m\mu}{2\hbar} q\right)} \tag{B.24}$$

$$C = 1 \tag{B.25}$$

where $H_{k(x)}$ represents the $k$-th Hermite polynomial, given that, for $C = 1$, (B.23) is the recurrence formula. By introducing (B.24-25) in (B.22) it follows both that

$$E_k = (k + \frac{1}{2})\hbar\mu \ . \tag{B.26}$$

and

$$\begin{aligned} V_{qu_k} &= -\frac{m\mu^2}{2} q^2 + \left[ k \left( \frac{\frac{m\mu}{\hbar} q H_{k-1} - 2(n-1) H_{k-2}}{H_k} \right) + \frac{1}{2} \right] \hbar\mu \\ &= -\frac{m\mu^2}{2} q^2 + (k + \frac{1}{2})\hbar\mu \end{aligned} \tag{B.27}$$

It is worth mentioning that by (B.27) the stationary conditions of eigenstates

$$\dot{p}_i = -\frac{\partial (V_{(q)} + V_{qu})}{\partial q_i} = -\frac{\partial (k + \frac{1}{2})\hbar\mu}{\partial q_i} = 0 \, , \tag{B.28}$$

induced by the quantum potential make the eigenstates satisfying the irrotational quantum condition (B.7).

# Appendix C

## The mass density distribution of the eigenstates of the harmoic oscillator in presence of noise

In order to obtain the mass density distribution, for the generic $j$-th eigenstates of the harmonic oscillator in presence of noise , we use the expression

$$n^{1/2}{}_j = \bar{C}_j H_{j\left(q\sqrt{\frac{m\mu}{\hbar}}\right)} exp\left[ -\frac{1}{2}\frac{m\mu}{\hbar}\left(1 - g_{j(q)} \frac{2\sim kT}{\hbar\mu}\right) q^2 \right]. \tag{C.1}$$

where the explicit expression of $g_{j(q)}$ determines the solution.

Given that in the deterministic limit of quantum mechanics (i.e., $T \to 0$, $D \to 0$ and $\sim = 0$ ) the mass distribution density of the quantum harmonic oscillator (see Appendix B)

$$n_j^{1/2} = |\cdot_j| = C_j H_{j\left(\sqrt{\frac{m\mu}{\hbar}}q\right)} exp\left[-\frac{m\mu}{2\hbar}q^2\right] \qquad \text{(C.2)}$$

satisfy the equation

$$\frac{\hbar^2}{4m^2}\frac{\partial\left(\left(\frac{\partial^2}{\partial q^2}\left(2\ln H_j - \frac{q^2}{\Delta q_0{}^2}\right)\right)+\frac{1}{2}\left(\frac{\partial}{\partial q}\left(2\ln H_j - \frac{q^2}{\Delta q_0{}^2}\right)\right)^2\right)}{\partial q} = \mu^2 q \qquad \text{(C.3)}$$

where $H_j$ are the Hermite polynomials and where the variance $\Delta q_0{}^2 = \left(\frac{\hbar}{m\mu}\right)$, it follows that

$$\frac{\hbar^2}{4m^2}\frac{\partial\left(\begin{array}{l}\left(\frac{\partial^2}{\partial q^2}\left(2\ln H_j - \frac{q^2}{\Delta q_0^2}\left(1-g_{j(q)}\frac{2\sim kT}{\hbar\mu}\right)\right)\right)\\ +\frac{1}{2}\left(\frac{\partial}{\partial q}\left(2\ln H_j - \frac{q^2}{\Delta q_0^2}\right)+\frac{\partial}{\partial q}\frac{q^2}{\Delta q_0^2}g_{j(q)}\frac{2\sim kT}{\hbar\mu}\right)^2\end{array}\right)}{\partial q}, \qquad \text{(C.4)}$$

$$= \copyright D\frac{\partial\left(2\ln H_j - \frac{q^2}{\Delta q_0^2}\left(1-g_{j(q)}\frac{2\sim kT}{\hbar\mu}\right)\right)}{\partial q} + \mu^2 q$$

that

$$\frac{\partial\left(\left(\frac{\partial^2}{\partial q^2}\frac{q^2}{\Delta q_0^2}g_{j(q)}\right)+\frac{1}{2}\left(\frac{2\sim mkT}{\hbar^2}\right)\left(\frac{\partial}{\partial q}\frac{q^2}{\Delta q_0^2}g_{j(q)}\right)^2+\left(\frac{\partial}{\partial q}\frac{q^2}{\Delta q_0^2}g_{j(q)}\right)\frac{\partial}{\partial q}\left(2\ln H_j - \frac{q^2}{\Delta q_0^2}\right)\right)}{\partial q}$$

$$= \frac{\hbar^2}{2\sim mkT}\frac{\partial\left(2\ln H_j + \frac{q^2}{\Delta q_0^2}g_{j(q)}\frac{2\sim kT}{\hbar\mu} - \frac{q^2}{\Delta q_0{}^2}\right)}{\partial q}$$

(C.5)

that, by posing $g_{j(q)} = \frac{1}{2} + \frac{\Delta q_0{}^2}{q^2} f_{j(q)}$ leads to

$$\frac{\partial\left(\begin{array}{l}\left(\frac{\partial^2}{\partial q^2} f_{j(q)}\right)+\frac{1}{2}\left(\frac{2\sim kT}{\hbar\mu}\right)\left(\frac{2q}{\Delta q_0^2}+\frac{\partial}{\partial q} f_{j(q)}\right)^2 \\ +\left(\frac{\partial}{\partial q}\left(\frac{q^2}{2\Delta q_0^2}+f_{j(q)}\right)\right)\frac{\partial}{\partial q}\left(2\ln H_j-\frac{q^2}{\Delta q_0^2}\right)\end{array}\right)}{\partial q}$$

$$=\frac{4m\mu}{\hbar}\frac{\partial\left(2\ln H_j+\left(\frac{q^2}{2\Delta q_0^2}+f_{j(q)}\right)\frac{2\sim kT}{\hbar\mu}-\frac{q^2}{\Delta q_0^2}\right)}{\partial q} \qquad \text{(C.6)}$$

that in the small noise assumption (i.e., $\frac{2kT}{\hbar\mu} \ll 1$) and $\sim \ll 1$, (C.6) can be approximated to

$$\frac{\partial\left(\left(\frac{\partial^2}{\partial q^2} f_{j(q)}\right)+\left(\frac{1}{2}\frac{\partial}{\partial q}\frac{q^2}{\Delta q_0^2}+\frac{\partial}{\partial q} f_{j(q)}\right)\frac{\partial}{\partial q}\left(2\ln H_j-\frac{q^2}{\Delta q_0^2}\right)\right)}{\partial q}$$

$$\cong\frac{4m\mu}{\hbar}\frac{\partial\left(2\ln H_j-\frac{q^2}{2\Delta q_0^2}\right)}{\partial q}=4\frac{\partial\left(\frac{2\ln H_j}{\Delta q_0^2}-\frac{1}{2}\left(\frac{q}{\Delta q_0^2}\right)^2\right)}{\partial q} \qquad \text{(C.7)}$$

and to

$$\frac{\partial\left(\left(\frac{\partial^2}{\partial q^2} f_{j(q)}\right)+\left(\frac{\partial}{\partial q} f_{j(q)}\right)\frac{\partial}{\partial q}\left(2\ln H_j-\frac{q^2}{\Delta q_0^2}\right)\right)}{\partial q}$$

$$\cong\frac{\partial\left(\frac{8\ln H_j}{\Delta q_0^2}-2\left(\frac{q}{\Delta q_0^2}\right)^2-\left(\frac{q}{\Delta q_0^2}\right)\left(2\frac{\partial}{\partial q}\ln H_j-2\frac{q}{\Delta q_0^2}\right)\right)}{\partial q} \qquad \text{(C.8)}$$

$$\cong\frac{\partial\left(\frac{8\ln H_j}{\Delta q_0^2}-\left(\frac{2q}{\Delta q_0^2}\frac{\partial}{\partial q}\ln H_j\right)\right)}{\partial q}$$

$$\left(\frac{\partial^2}{\partial q^2} f_{j(q)}\right)+\left(\frac{\partial}{\partial q} f_{j(q)}\right)\left(2\frac{\partial}{\partial q} ln H_j - 2\frac{q}{\Delta q_0^2}\right)$$

$$\cong \frac{8\, ln H_j}{\Delta q_0^{\;2}} - \left(\frac{2q}{\Delta q_0^2}\frac{\partial}{\partial q} ln H_j\right) + C \tag{C.9}$$

and to

$$H_j\left(\frac{\partial^2}{\partial q^2} f_{j(q)}\right)+\left(\frac{\partial}{\partial q} f_{j(q)}\right)\left(4j\frac{H_{j-1}}{\Delta q_0} - 2q\frac{H_j}{\Delta q_0^2}\right) - \frac{8H_j\, ln H_j}{\Delta q_0^{\;2}} - \left(4jq\frac{H_{j-1}}{\Delta q_0^3}\right) = 0 \tag{C.10}$$

that is the Bernoulli equation for $f'_{j(q)} = \frac{\partial}{\partial q} f_{j(q)}$ whose solution reads

$$f'_{j(q)} = \frac{\int_{q_0}^{q}\left(\frac{8H_{j(x)}\, ln H_{j(x)}}{\Delta q_0^{\;2}} - \left(4jx\frac{H_{j-1(x)}}{\Delta q_0^3}\right)\right) exp[\int\left(4j\frac{H_{j-1(t)}}{\Delta q_0} - 2t\frac{H_{j(t)}}{\Delta q_0^2}\right)dt\,]\,dx}{exp[\int_{q_0}^{q}\left(4j\frac{H_{j-1(x)}}{\Delta q_0} - 2x\frac{H_{j(x)}}{\Delta q_0^2}\right)dx\,]} + f'_{j(q_0)} \tag{C.11}$$

Moreover, since for the symmetry of the system $f'_{j(q_0=0)} = 0$, it follows that

$$f_{j(q)} = \int_0^q \frac{\int_0^x\left(\frac{8H_{j(u)}\, ln H_{j(u)}}{\Delta q_0^{\;2}} - \left(4jx\frac{H_{j-1(u)}}{\Delta q_0^3}\right)\right) exp[\int_0^u\left(4j\frac{H_{j-1(t)}}{\Delta q_0} - 2t\frac{H_{j(t)}}{\Delta q_0^2}\right)dt\,]\,du}{exp[\int_0^x\left(4j\frac{H_{j-1(u)}}{\Delta q_0} - 2x\frac{H_{j(u)}}{\Delta q_0^2}\right)du\,]}dx + Const \tag{C.12}$$

Where the constant can be set to zero by the appropriate choice of the normalization coefficient $\bar{C}_j$. Finally, the mass density distribution of the *j*-th eigenstate in presence of noise reads

$$n^{1/2}{}_j = \bar{C}_j H_{j\left(q\sqrt{\frac{m\mu}{\hbar}}\right)}\, exp\left[-\frac{1}{2}\left(\frac{m\mu}{\hbar} - \left(1+\frac{\Delta q_0^{\;2}}{q^2} f_{j(q)}\right)\frac{2\sim mkT}{\hbar^2}\right)q^2\right]$$

$$\cong \bar{C}_j\, exp\left[-f_{j(q)}\frac{\sim kT}{\hbar\mu}\right] H_{j\left(q\sqrt{\frac{m\mu}{\hbar}}\right)}\, exp\left[-\frac{1}{2}\left(\frac{m\mu}{\hbar} - \frac{2\sim mkT}{\hbar^2}\right)q^2\right] \tag{C.13}$$

that for *j*=0 gives

$$\left(\frac{\partial^2}{\partial q^2} f_{0(q)}\right) - \left(2\frac{q}{\Delta q_0^2}\frac{\partial}{\partial q} f_{0(q)}\right) = 0\,, \tag{C.14}$$

that is satisfacted for $f_{0(q)} = H_{0\left(q\sqrt{\frac{m\mu}{\hbar}}\right)} = 1$, so that formula (42) is confirmed, given that

$$n^{1/2}{}_{0} = \bar{C}_0\, exp\left[-f_{0(q)}\frac{\sim kT}{\hbar\mu}\right] H_{0\left(q\sqrt{\frac{m\mu}{\hbar}}\right)}\, exp\left[-\frac{1}{2}\left(\frac{m\mu}{\hbar} - \frac{2\sim mkT}{\hbar^2}\right)q^2\right]$$

$$\cong \bar{C}_0\left(1 - \frac{\sim kT}{\hbar\mu}\right) exp\left[-\frac{1}{2}\left(\frac{m\mu}{\hbar} - \frac{2\sim mkT}{\hbar^2}\right)q^2\right] \qquad \text{(C.15)}$$

$$\cong \bar{C}_0\, exp\left[-\frac{1}{2}\left(\frac{m\mu}{\hbar} - \frac{2\sim mkT}{\hbar^2}\right)q^2\right]$$

# Appendix D

## The quantum-stochastic PTF

By using the identities

$$< q_{k+1} >= q_k + < \dot{q}_k > \Delta t_n + < \ddot{q}_k > \frac{\Delta t_n{}^2}{2}$$

$$= q_k - \frac{1}{m©}\frac{\partial\left(V_{(q_k)} - (\frac{\hbar^2}{4m})\left(\left(\frac{\partial^2 ln\, n_{(\bar{q}_k)}}{\partial q^2}\right) - \frac{1}{2}\left(\frac{\partial ln\, n_{(\bar{q}_k)}}{\partial q}\right)^2\right)\right)}{\partial q_k}\Delta t_n \quad \text{(D.1)}$$

$$-\frac{1}{m©}\frac{d}{dt}\frac{\partial\left(V_{(q_k)} - (\frac{\hbar^2}{4m})\left(\left(\frac{\partial^2 ln\, n_{(\bar{q}_k)}}{\partial q^2}\right) - \frac{1}{2}\left(\frac{\partial ln\, n_{(\bar{q}_k)}}{\partial q}\right)^2\right)\right)}{\partial q_k}\frac{\Delta t_n{}^2}{2}$$

that with the mid-point approximation $q_k \to \bar{q}_k = \frac{q_{k+1} + q_k}{2}$, $< \dot{q}_k > \to < \dot{\bar{q}}_k >$ and $< \ddot{q}_k > \to < \ddot{\bar{q}}_k >$, reads

$$< q_{k+1} >= q_k - \frac{1}{m©}\frac{\partial\left(V_{(q_k)} - (\frac{\hbar^2}{4m})\left(\left(\frac{\partial^2 ln\, n_{(\bar{q}_k)}}{\partial q^2}\right) - \frac{1}{2}\left(\frac{\partial ln\, n_{(\bar{q}_k)}}{\partial q}\right)^2\right)\right)}{\partial q_k}\Delta t_n$$

$$-\frac{1}{m©}\frac{d}{dt}\frac{\partial\left(V_{(q_k)} - (\frac{\hbar^2}{4m})\left(\left(\frac{\partial^2 ln\, n_{(\bar{q}_k)}}{\partial q^2}\right) - \frac{1}{2}\left(\frac{\partial ln\, n_{(\bar{q}_k)}}{\partial q}\right)^2\right)\right)}{\partial q_k}\frac{\Delta t_n{}^2}{2} \quad \text{(D.2)}$$

in the noise probability function, we obtain

$$lim_{\Delta t\to 0}\,\mathcal{P}(\Delta\mathcal{W}_k,\Delta t)=\left(4\text{®}\,D\Delta t\right)^{-1/2}exp-\frac{\Delta\mathcal{W}_k^{\;2}}{4\Delta t}=\left(4\text{®}\,D\Delta t\right)^{-1/2}exp-\frac{1}{4\Delta t}\frac{\left(q_{k+1}-<q_{k+1}>\right)^2}{D}$$

$$=\left(4\text{®}\,D\Delta t\right)^{-1/2}exp-\frac{1}{4\Delta t}\frac{\left(q_{k+1}-q_k-<\dot{\bar{q}}_k>\Delta t-\frac{<\ddot{\bar{q}}_k>}{2}\Delta t^2\right)^2}{D}$$

(D.3)

where

$$<\dot{\bar{q}}_k>=-\frac{1}{m\text{©}}\frac{\partial\left(V_{(\bar{q}_k)}-(\frac{\hbar^2}{4m})\left(\left(\frac{\partial^2\,ln\,n_{(\frac{q_{k+1}+q_k}{2})}}{\partial q^2}\right)-\frac{1}{2}\left(\frac{\partial\,ln\,n_{(\frac{q_{k+1}+q_k}{2})}}{\partial q}\right)^2\right)\right)}{\partial\bar{q}_k} \qquad (D.4)$$

and

$$<\ddot{\bar{q}}_k>=-\frac{1}{m\text{©}}\frac{d}{dt}\frac{\partial\left(V_{(\bar{q}_k)}-(\frac{\hbar^2}{4m})\left(\left(\frac{\partial^2\,ln\,n_{(\bar{q}_k)}}{\partial q^2}\right)-\frac{1}{2}\left(\frac{\partial\,ln\,n_{(\bar{q}_k)}}{\partial q}\right)^2\right)\right)}{\partial\bar{q}_k} \qquad (D.5)$$

that, by applying the conservation of the PMD that in the discrete time reads

$$n(q_{k+1},(k+1)\Delta t)=\int_{-\infty}^{\infty}\mathcal{P}(q_{k+1},q_k(|\Delta t,k\Delta t)n(q_k,k\Delta t)dq_k \qquad , \qquad (D.6)$$

allows to obtain $n(q_{k+1},(k+1)\Delta t)$ from $n(q_k,k\Delta t)$. By comparing (D.6) with (61) we can see that in the small fluctuation limit it follows that

$$lim_{\Delta t\to 0}\mathcal{P}(q_{k+1},q_k\,|\,\Delta t,k\Delta t)=P(q_{k+1},q_k\,|\,\Delta t,k\Delta t)\;. \qquad (D.7)$$

Given the initial condition $n_{(q,0)}$, with the help of (D.7) it is possible to obtain the solution in the discrete form that reads

$$\begin{aligned}
&P(q_n,q_0,t,t_0)\equiv P(q_n,q_0\,|\,t-t_0,0)=lim_{\Delta t\to 0}\,\mathcal{P}(q_n,q_0\,|\,n\Delta t,0)\\
&=lim_{\Delta t\to 0}\int\mathcal{P}(q_n,q_{n-1}\,|\,\Delta t,(n-1)\Delta t)P(q_{n-1},q_{n-2}\,|\,\Delta t,(n-2)\Delta t)dq_{n-1}\\
&=lim_{\Delta t\to 0}\int\mathcal{P}(q_n,q_{n-1}\,|\,\Delta t,(n-1)\Delta t)P(q_{n-1},q_{n-2}\,|\,\Delta t,(n-2)dq_{n-1}\ldots \qquad (D.8)\\
&\qquad\ldots\int\mathcal{P}(q_2,q_1\,|\,\Delta t,(n-(n-1))\Delta t)dq_2\int\mathcal{P}(q_1,q_0\,|\,\Delta t,0)dq_1\\
&=lim_{\Delta t\to 0}\int_{-\infty}^{\infty}\prod_{k=1}^{n}dq_{k-1}\mathcal{P}(q_k,q_{k-1}\,|\,\Delta t,(k-1)\Delta t)
\end{aligned}$$

that by using (D.3) leads to

$$P(q_n, q_0 \mid t-t_0, t_0) =$$

$$= lim_{\Delta t\to 0}\left\{\prod_{k=1}^{n}\int_{-\infty}^{\infty} dq_{k-1}\right\}\left(4\text{®}D\Delta t\right)^{-n/2} exp-\frac{\Delta t}{4D}\sum_{k=1}^{n}\left(\frac{q_k - q_{k-1} - <\dot{q}_{k-1}>\Delta t - \frac{<\ddot{q}_{k-1}>}{2}\Delta t^2}{\Delta t}\right)^2$$

$$= lim_{\Delta t\to 0}\left\{\prod_{k=1}^{n}\int_{-\infty}^{\infty} dq_{k-1}\right\}\left(4\text{®}D\Delta t\right)^{-n/2} exp-\frac{\Delta t}{4D}\sum_{k=1}^{n}\left(\frac{q_k - q_{k-1}}{\Delta t} - <\dot{q}_{k-1}> - \frac{<\ddot{q}_{k-1}>}{2}\Delta t\right)^2$$

$$\cong lim_{\Delta t\to 0}\left\{\prod_{k=1}^{n}\int_{-\infty}^{\infty} dq_{k-1}\right\}\left(4\text{®}D\Delta t\right)^{-n/2} exp-\frac{\Delta t}{4D}\sum_{k=1}^{n}\begin{bmatrix}\left(\frac{q_k - q_{k-1}}{\Delta t} - <\dot{q}_{k-1}>\right)^2 \\ -D^{1/2}\frac{\Delta \mathcal{W}_n}{\Delta t}<\ddot{q}_{k-1}>\Delta t\end{bmatrix}$$

$$= lim_{\Delta t\to 0}\left\{\prod_{k=1}^{n}\int_{-\infty}^{\infty} dq_{k-1}\right\}\left(4\text{®}D\Delta t\right)^{-n/2} exp-\frac{\Delta t}{4D}\begin{bmatrix}\sum_{k=1}^{n}\left(\frac{q_k - q_{k-1}}{\Delta t}\right)^2 + <\dot{q}_{k-1}>^2 \\ +2\left(\frac{q_k - q_{k-1}}{\Delta t}\right)<\dot{q}_{k-1}> \\ -\sum_{k=1}^{n}D^{1/2}\frac{\Delta \mathcal{W}_k}{\Delta t}<\ddot{q}_{k-1}>\Delta t\end{bmatrix} \tag{D.9}$$

Moreover, since passing to the continuous limit the following identity holds

$$\begin{aligned}
&lim_{\Delta t\to 0} D^{1/2}\frac{\Delta \mathcal{W}_k}{\Delta t}<\ddot{q}_{k-1}>_{(q_{k-1})}\Delta t = lim_{\Delta t\to 0} D^{1/2}\frac{\Delta \mathcal{W}_k}{\Delta t}\frac{\partial <\dot{q}_{k-1}>}{\partial q}\dot{q}_{(q_{k-1})}\Delta t \\
&= lim_{\Delta t\to 0} D^{1/2}\, 2 < \frac{\Delta \mathcal{W}_k}{\Delta t}\dot{q}_{(q_{k-1})}, \frac{\Delta \mathcal{W}_k}{\Delta t}\dot{q}_{(q_{k-1})} >_{\Delta t}^{1/2}\frac{\partial <\dot{q}_{k-1}>}{\partial q}\Delta t \\
&= lim_{\Delta t\to 0} D^{1/2}\, 2 < \text{¶}_{(t)}, \text{¶}_{(t)} >_{\Delta t}^{1/2} < \dot{q}_{(q_{k-1})} >_{\Delta t}\frac{\partial <\dot{q}_{k-1}>}{\partial q}\Delta t \\
&= lim_{\Delta t\to 0} D^{1/2}\, 2\sqrt{2\Delta t}\,\frac{\partial <\dot{q}_{k-1}>}{\partial q} < \dot{q}_{(q_{k-1})} > \Delta t
\end{aligned} \tag{D.10}$$

where

$$\begin{aligned}
&= D^{1/2}2\sqrt{2\Delta t}\,\frac{\partial <\dot{\overline{q}}_{k-1}>}{\partial \overline{q}}<\dot{\overline{q}}_{k-1}>\Delta t = D^{1/2}2\sqrt{2\Delta t}\,\frac{\partial <\dot{\overline{q}}_{k-1}>}{\partial \overline{q}}\frac{d}{dt}<\overline{q}_{k-1}>\Delta t \\
&= D^{1/2}2\sqrt{2\Delta t}\,\frac{\partial <\dot{\overline{q}}_{k-1}>}{\partial \overline{q}}\frac{\Delta <\overline{q}_{k-1}>}{\Delta t}\Delta t = D^{1/2}2\sqrt{2\Delta t}\,\frac{\partial <\dot{\overline{q}}_{k-1}>}{\partial \overline{q}}\frac{<\overline{q}_{k-1}>_{t_k} - <\overline{q}_{k-1}>_{t_{k-1}}}{\Delta t}\Delta t \\
&= D^{1/2}2\sqrt{2\Delta t}\,\frac{\partial <\dot{\overline{q}}_{k-1}>}{\partial \overline{q}}\frac{\overline{q}_{k-1} + <\Delta\overline{q}_{k-1}, \Delta\overline{q}_{k-1}>_{\Delta t}^{1/2} - \overline{q}_{k-1}}{\Delta t}\Delta t \\
&= D^{1/2}2\sqrt{2\Delta t}\,\frac{\partial <\dot{\overline{q}}_{k-1}>}{\partial \overline{q}}\frac{d\sqrt{2D\Delta t}}{d\Delta t}\Delta t \\
&= D4\sqrt{\Delta t}\,\frac{\partial <\dot{\overline{q}}_{k-1}>}{\partial \overline{q}}\frac{d\sqrt{\Delta t}}{d\Delta t}\Delta t = 2D\frac{\partial <\dot{\overline{q}}_{k-1}>}{\partial \overline{q}_{k-1}}\Delta t
\end{aligned}$$

(D.11)

finally, it follows that

$$P(q,q_0,t,t_0)=$$

$$=\lim_{\Delta t\to 0}\left\{\prod_{k=1}^{n}\int_{-\infty}^{\infty}dq_{k-1}\right\}\left(4®D\Delta t\right)^{-n/2}exp\left[\begin{array}{l}-\frac{1}{2D}\sum_{k=1}^{n}<\dot{\bar{q}}_{k-1}>\Delta q_k-\frac{\Delta t}{4D}\sum_{k=1}^{n}\left(\frac{q_k-q_{k-1}}{\Delta t}\right)^2+<\dot{\bar{q}}_{k-1}>^2\\ -\frac{\Delta t}{4D}\sum_{k=1}^{n}D^{1/2}\Delta\mathcal{W}_k'<\ddot{\bar{q}}_{k-1}>_{(q_{k-1})}\end{array}\right] \quad \text{(D.12)}$$

$$=\int_{q_0}^{q}\mathcal{D}q\,exp\left[-\frac{1}{2D}\sum_{k=1}^{n}<\dot{\bar{q}}_{k-1}>\Delta q_k\right]exp-\frac{\Delta t}{4D}\left[\sum_{k=1}^{n}\begin{array}{l}\left(\frac{q_k-q_{k-1}}{\Delta t}\right)^2+<\dot{\bar{q}}_{k-1}>^2\\ -2D\frac{\partial<\dot{\bar{q}}_{k-1}>}{\partial\bar{q}_{k-1}}\end{array}\right]$$